\documentclass[aps,pra,reprint,groupedaddress]{revtex4-2}
\usepackage{graphicx}
\usepackage{epstopdf}
\graphicspath{figures/}
\usepackage{dcolumn}
\usepackage{amssymb}
\usepackage{amsmath}
\usepackage{bm}
\usepackage{times}
\usepackage{mhchem}
\usepackage{lineno}
\usepackage[colorlinks,linkcolor=blue,anchorcolor=blue,citecolor=blue,urlcolor=blue]{hyperref}
\UseRawInputEncoding
\begin{document}
\title{Tripartite hybrid quantum systems: Skyrmion-mediated quantum interactions between single NV centers and superconducting qubits}
\author{Xue-Feng Pan}
\author{Peng-Bo Li}
\email{lipengbo@mail.xjtu.edu.cn}
\affiliation{Ministry of Education Key Laboratory for Nonequilibrium Synthesis and Modulation of Condensed Matter, Shaanxi Province Key Laboratory of Quantum Information and Quantum Optoelectronic Devices, School of Physics, Xi'an Jiaotong University, Xi'an 710049, China}

\date{\today}
\begin{abstract}
Nitrogen-vacancy (NV) centers in diamond and superconducting qubits are two promising solid-state quantum systems for quantum science and technology, but the realization of controlled interfaces between individual solid-state spins and superconducting qubits remains  fundamentally  challenging. Here, we propose and analyze a hybrid quantum system consisting of a magnetic skyrmion, an NV center and a superconducting qubit, where the solid-state qubits are both positioned in proximity to the skyrmion structure in a thin magnetic disk. We show that it is experimentally feasible to achieve strong magnetic (coherent or dissipative) coupling between the NV center and the superconducting qubit by using the \textit{quantized gyration mode of the skyrmion} as an intermediary. This allows coherent information
transfer and nonreciprocal responses between the NV center and the superconducting qubit at the single quantum level with high controllability.  The proposed platform provides a scalable pathway for implementing quantum protocols that synergistically exploit the complementary advantages of spin-based quantum memories, microwave-frequency superconducting circuits, and topologically protected magnetic excitations.
\end{abstract}

\maketitle

\section{\label{Int}Introduction}
Quantum information processing demands high-quality quantum systems with extended coherence times and external controllability.
Typical solid-state quantum systems for quantum technology include electronic spins in diamond~\cite{2013BarGillP17431743,2013DohertyP145,2016SipahigilP847850,2018SohnP20122012,2019BradacP56255625,2021NeumanP121121,2024HarrisP8541485414,2011FuchsP789793} and superconducting qubits~\cite{2013XiangP623653,2020KjaergaardP369395,2020ClerkP257267,2020BurkardP129140,2019AruteP505510,2020KjaergaardP369395,2023PitaVidalP11101115}.
Single NV centers serve as superior quantum memory platforms owing to their exceptional coherence lifetimes and immunity to inhomogeneous broadening inherent in spin ensembles~\cite{2011AharonovichP397405,2013BarGillP17431743,2013DohertyP145,2013MacQuarrieP227602227602,2014OvartchaiyapongP44294429,2014TeissierP2050320503,2016GolterP4106041060,2016GolterP143602143602,2020ChenP82678272,2020BarryP1500415004,2023OrphalKobinP1104211042}.
In contrast, superconducting qubits stand out as high-performance quantum processors due to their unparalleled gate fidelity and microwave control precision~\cite{2020KjaergaardP369395,2021BlaisP2500525005,2020BurkardP129140,2008ClarkeP10311042,2017WendinP106001106001}.
Therefore, it is quite appealing to integrate these two kinds of solid-state quantum systems in a single hybrid quantum setup~\cite{2013XiangP623653,2020ClerkP257267,2020KjaergaardP369395,2020BurkardP129140}, which would allow the realization of long-lived quantum memories for superconducting qubits.
However, the direct magnetic coupling between a single spin and a single superconducting qubit is typically only a few hertz, much smaller than the relevant decoherence rates~\cite{2013XiangP623653,2010MarcosP210501210501}.
This remains a fundamental barrier in constructing NV-superconducting qubit hybrid architectures.

Recent theoretical and experimental proposals demonstrate that magnonic modes in micromagnet spheres~\cite{2024BinP4360143601,2020ColombanoP147201147201,2020GonzalezBallesteroP9360293602,2020GonzalezBallesteroP125404125404,2018HarderP137203137203,2023HeiP7360273602,2022KaniP1360213602,2022KaniP257201257201,2019LachanceQuirionP7010170101,2018LiP203601203601,2021LiP4034440344,2018PirmoradianP224409224409,2023QianP34373437,2022ShenP243601243601,2022ShenP123601123601,2014TabuchiP8360383603,2015TabuchiP405408,2018WangP5720257202,2022WangP75807580,2020WolskiP117701117701,2020YangP147202147202,2024YangP206902206902,2020YuanP5360253602,2020YuP107202107202,2014ZhangP156401156401,2015ZhangP1501415014,2016ZhangP15012861501286,2022BittencourtP183603183603,2010SoykalP7720277202,2022KounalakisP3720537205,2023HeiP7360273602,2021HeiP4370643706,2023JiP180409180409,2020NeumanP247702247702,2022XiongP245310245310} can achieve strong coupling at the single quantum level ($\sim\mathrm{MHz}$ scale) to both NV centers and superconducting qubits, offering a potential quantum interface between these solid-state qubits~\cite{2020NeumanP247702247702,2021HeiP4370643706,2022XiongP245310245310,2023JiP180409180409,2023HeiP7360273602}.
This coupling mechanism could bridge individual spin defects with macroscopic quantum circuits through magnon-mediated interactions.
Nevertheless, a fundamental scaling dichotomy arises: NV centers necessitate sub-100-nm-scale magnetic confinement (radius $<$100 nm) for optimal coupling enhancement~\cite{2020NeumanP247702247702,2021HeiP4370643706}, while superconducting qubits (SQ) require micron-scale magnetic structures (radius $>$1 $\mu$m) to maximize interaction strength~\cite{2022KounalakisP3720537205}.
This imposes mutually exclusive scaling constraints on the hybrid system design.
In addition, a concurrent integration challenge emerges from the inherent incompatibility between three-dimensional micromagnet geometries and planar superconducting resonator architectures, where sub-micrometer alignment precision exceeds the practical limits of conventional flip-chip bonding technique.

To overcome the above mentioned challenges, we here propose to exploit skyrmion structures in thin magnetic disks ~\cite{2013NagaosaP899911,2017FertP1703117031,2019OchoaP19300051930005,2022ReichhardtP3500535005,2019HirschbergerP58315831,2020KhanhP444449,2021YuP50795079,2022SekiP181187,2024HirschbergerP4545,2015RommingP177203177203,2019MeyerP38233823,2019CortesOrtunoP214408214408,2022BrueningP241401241401,2015ZhouP81938193,2016ZhangP1029310293,2021GoebelP128,2014SchuetteP9442394423,2021WangP3720237202,2023JinP166704166704,2013IwasakiP14631463,2015ZhangP102401102401,2019LiuP1400414004,2020YaoP8303283032,2023WangP9440494404,2023OkumuraP6670266702,2017OchoaP2041020410,2015HagemeisterP84558455,2023KatoP54165416,2015ZhangP1136911369,2018LuoP11801184,2006RoesslerP797801,2010ButenkoP5240352403,2014BanerjeeP3104531045,2014JansonP53765376,2014WilsonP9441194411,2015LeonovP82758275,2016LeonovP6500365003,2016LinP6443064430,2016TakashimaP134415134415,2018WangP3131,2019LohaniP4106341063} for a quantum interface between NV centers and superconducting qubits.
Recent studies have began to investigate  the construction of qubits using the quantized helicity degree of freedom in skyrmions and their interactions to other quantum systems~\cite{2021PsaroudakiP6720167201,2023XiaP106701106701,2024PanP193601193601,2024PanP2306723067}. In addition to the helicity degree of freedom, a magnetic skyrmion texture also exhibits a distinct excitation mode known as the skyrmion gyration mode (GM)~\cite{2016MruczkiewiczP174429174429,2018GareevaP3500935009,2020LiuP166965166965,2023LiuP170649170649}.
Compared to magnon modes in micromagnetic spheres, the skyrmion gyration mode in a thin disk has its own distinct features. First, the stray field profiles of the gyration mode are concentrated near the skyrmion core, where the exponentially localized magnetic fields can significantly enhance the coupling to other quantum systems.
Second, the confinement of the skyrmion gyration mode within a two-dimensional disk makes it an ideal candidate for integration with other quantum systems on a chip.
It is therefore highly appealing to employ the quantized gyration modes of skyrmions as a quantum interface for solid-state qubits.

We consider a tripartite hybrid quantum system consisting of a magnetic skyrmion, an NV center, and a flux-tunable transmon qubit.
Stabilized skyrmions have been extensively studied, with their observation and control demonstrated in materials like \ce{MnSi_{1-x}Ge_x}, \ce{VOSe_2O_5}, \ce{Cu_2OSeO_3}, \ce{GaV_4S8}, etcd~\cite{2009MuehlbauerP915919,2010MuenzerP4120341203,2010YuP901904,2011HeinzeP713718,2015KezsmarkiP11161122,2017NayakP561566,2018WooP288296,2019KurumajiP914918,2021TokuraP28572897,2024PhamP307312,2017HrabecP1576515765,2019CasiraghiP145145,2023DaiP68366836,2018ZhangP21152115,2017WooP1557315573,2020PengP181186,2013RommingP636639,2015HannekenP10391042,2017HsuP123126,2019PeriniP237205237205}.
The NV center, positioned above the skyrmion, is coupled to the stray field of the gyration mode through the magnetic dipole coupling.
Below the magnetic disk, a flux-tuned transmon qubit is positioned.
The magnetic field generated by the disk passes through the superconducting quantum interference device (SQUID), producing a magnetic flux that enables strong coupling between the skyrmion and the transmon qubit. These types of coupling differ fundamentally from the previously studied NV-phonon~\cite{2009RablP4130241302,2013BennettP156402156402,2016GolterP4106041060,2020LiP153602153602,2023PanP2372223722,2024FungP263602263602}, NV-magnon~\cite{2020NeumanP247702247702,2021HeiP4370643706,2023HeiP7360273602}, skyrmion-magnon~\cite{2024PanP193601193601}, and SQ-magnon~\cite{2015TabuchiP405408,2022KounalakisP3720537205} couplings.
We show that by making use of the virtual excitations of the  gyration mode, strong magnetic (coherent or dissipative) coupling between the NV center and the transmon qubit can be achieved, with a coupling strength of MHz. This allows quantum information transfer and nonreciprocal response between the two solid-state qubits.
This work provides a promising quantum platform for on-chip  quantum information processing with purely solid-state
few-body hybrid quantum systems.

The structure of this paper is as follows.
Sec.~\ref{CPSGM} presents a comprehensive analysis of the classical dynamical behavior of the skyrmion gyration modes, while Sec.~\ref{QPSGM} explores their quantization properties.
Secs.~\ref{SNHQS} and \ref{IBST} delve into the detailed coupling mechanisms between the skyrmion gyration mode and two key quantum components: NV centers and transmon qubits.
Sec.~\ref{ICBNSQ} introduces an indirect coupling scheme between NV centers and transmon qubits mediated by the skyrmion gyration mode.
Sec.~\ref{EF} demonstrates the experimental feasibility of the proposed theoretical model through multidimensional analysis.
Finally, Sec.~\ref{Con} concludes the paper.

\section{\label{CPSGM}Classical properties of the skyrmion gyration mode}
\subsection{\label{EoS}The Stationary Skyrmion}
Skyrmions are topological solitons in magnetic materials characterized by a centrosymmetric spiral structure.
The stabilization of skyrmions arises from the competition among various interactions in magnetic materials, including exchange interactions, Dzyaloshinskii-Moriya (DM) interactions, easy-axis anisotropy, and Zeeman interactions.
Materials with DM interactions, known as chiral materials, exhibit skyrmions that have been theoretically predicted and experimentally observed.
The energy density functional of skyrmions can be expressed in terms of the normalized magnetization $\boldsymbol{m}=\boldsymbol{M}/M_S$~\cite{2018PsaroudakiP237203237203,2020PsaroudakiP9720297202}
\begin{equation}
	\mathcal{F}_{\rm{Sk}}=A_{\rm{ex}}\left(\nabla\boldsymbol{m}\right)^2+\mathcal{F}_D-K \left(\boldsymbol{m}\cdot \boldsymbol{u}\right)^2-\mu_0 M_S\boldsymbol{m}\cdot \boldsymbol{H}_e,
\end{equation}
where $M_S$ represents the saturation magnetization of the magnetic material.
The first term represents the energy density contributed by the isotropic exchange interaction, with $A_{\rm{ex}}$ being the exchange stiffness constant.
The second term accounts for the energy density contributed by the DM interaction, which includes two types: bulk DM interaction and interface DM interaction, denoted as
$\mathcal{F}_D^b=D\boldsymbol{m}\cdot(\nabla\times\boldsymbol{m})$ and $\mathcal{F}_D^i=D[m_z(\nabla\cdot\boldsymbol{m})-(\boldsymbol{m}\cdot \nabla)m_z]$, respectively, with $D$ being the DM constant.
The former stabilizes Bloch skyrmions, while the latter stabilizes N\'eel-type skyrmions.
The third term corresponds to the uniaxial anisotropy perpendicular to the surface, with $K$ being the anisotropy constant and $\boldsymbol{u}$ denoting the anisotropy axis.
The final term is the Zeeman energy density caused by the external magnetic field $\boldsymbol{H}_e$.
The energy of the skyrmion can be expressed as the integral of its energy density $\mathcal{H}_{\rm{Sk}}=\int d\boldsymbol{r} \mathcal{F}_{\rm{Sk}}$.
The dynamics of the skyrmion is described by the Landau-Lifshitz-Gilbert (LLG) equation~\cite{2004GilbertP34433449}:
\begin{equation}
	\frac{d\boldsymbol{m}}{dt}=-\gamma_e\boldsymbol{m}\times\boldsymbol{H}_{\rm{eff}}+\alpha\boldsymbol{m}\times\frac{d\boldsymbol{m}}{dt},
	\label{LLG}
\end{equation}
with the effect field $\boldsymbol{H}_{\rm{eff}}=-[1/(\mu_0 M_S)]\delta \mathcal{H}_{\rm{Sk}}/\delta \boldsymbol{m}$, the Gilbert damping coefficient $\alpha$, and the gyromagnetic ratio $\gamma_e$.

Since the LLG equation is nonlinear, obtaining an analytical solution for the skyrmion's magnetization field $\boldsymbol{m}$ is challenging.
Without loss of generality, we consider the Belavin-Polyakov (BP) model here, which is an approximation of the skyrmion~\cite{2017GuslienkoP176182,1975BelavinP245245}.
For convenience, we write the normalized magnetization vector as $\boldsymbol{m}=\left(m_x,m_y,m_z\right)$.
In the BP model, the components of the magnetization vector $\boldsymbol{m}$ are given by
\begin{subequations}
	\begin{align}
		m_x&=c\frac{2x \cos \Phi_0-2y\sin\Phi_0}{c^2+\left(x^2+y^2\right)}, \\
		m_y&=c\frac{2x \sin \Phi_0+2y\cos\Phi_0}{c^2+\left(x^2+y^2\right)}, \\
		m_z&=\mathcal{P}\frac{c^2-\left(x^2+y^2\right)}{c^2+\left(x^2+y^2\right)},
	\end{align}
	\label{BPmodel}
\end{subequations}
where $\mathcal{P}=\pm1$ represents the core polarity of the skyrmion and $\Phi_0$ denotes the skyrmion phase.
The symbol $c$ represents the reduced radius of the skyrmion, defined as $c=R_{\rm{Sk}}/R$, where $R_{\rm{Sk}}$ is the actual radius of the skyrmion and $R$ is the disk radius.
For Bloch skyrmions, $\Phi_0=\pm\pi/2$ and the skyrmion chirality is defined as $\mathcal{C}=\sin\Phi_0$; for N\'eel skyrmions, $\Phi_0=0, \pi$ and the skyrmion chirality is defined as $\mathcal{C}=\cos\Phi_0$.
Here, we focus on Bloch skyrmions, setting the chirality $\mathcal{C}=1$ and the polarity $\mathcal{P}=-1$.

\subsection{\label{ThieleEqation}The Thiele equation}
Here, we focus on the low-frequency dynamics of the skyrmion (gyrotropic eigenmodes), where it behaves as a rigid particle, preserving its internal structure during motion.
To characterize this motion, we introduce the collective coordinate of the skyrmion, specifically its center of mass $\boldsymbol{\mathcal{R}}_c=\left(X,Y\right)=\int \boldsymbol{r}\rho(\boldsymbol{r}) d\boldsymbol{r}/\int \rho(\boldsymbol{r}) d\boldsymbol{r}$, where $\rho(\boldsymbol{r})$ represents the topological charge density of the skyrmion.
Using the collective coordinate $\boldsymbol{\mathcal{R}}_c$, the magnetization configuration of the skyrmion at any given moment can be expressed as $\boldsymbol{m}(\boldsymbol{r},t)=\boldsymbol{m}[\boldsymbol{r}-\boldsymbol{\mathcal{R}}_c(t)]$.
Since the magnetization configuration $\boldsymbol{m}$ remains unchanged during motion, we can substitute $\boldsymbol{m}(\boldsymbol{r},t)=\boldsymbol{m}[\boldsymbol{r}-\boldsymbol{\mathcal{R}}_c(t)]$ into Eq.~(\ref{LLG}), and integrate over the skyrmion’s magnetization profile to derive the equation of motion for the collective coordinate $\boldsymbol{\mathcal{R}}_c$, which yields
\begin{equation}
	\dot{\boldsymbol{\mathcal{R}}}_c\times\boldsymbol{G}+\boldsymbol{F}=\bar{\bar{D}}\dot{\boldsymbol{\mathcal{R}}}_c.
	\label{ThieleEq}
\end{equation}
The gyrocoupling vector $\boldsymbol{G}$ is defined as $\boldsymbol{G}=(0,0,G)=4\pi h_G M_SQ/\gamma_e \boldsymbol{e}_z$ and related to the material's thickness $h_G$, saturation magnetization $M_S$, and topological charge $Q = (1/4\pi)\int d\boldsymbol{r} \left(\partial_x\boldsymbol{m}\times\partial_y\boldsymbol{m}\right)\cdot\boldsymbol{m}$.
The symbol $\bar{\bar{D}}$ is the dissipation tensor, with its matrix elements defined as $\bar{\bar{D}}_{i,j}=(\alpha h_G M_S/\gamma_e)\int d\boldsymbol{r} \partial_i\boldsymbol{m}\cdot \partial_j \boldsymbol{m}$, where $i,j=x,y$.
Notably, in contrast to magnetic quasiparticles like magnetic vortices, skyrmions exhibit a substantial inertial mass $\mathcal{M}$~\cite{2012MakhfudzP217201217201,2015BuettnerP225228}.
With the introduction of the skyrmion’s inertial mass $\mathcal{M}$, the equation of motion for the center of mass $\boldsymbol{\mathcal{R}}_c$ can be phenomenologically formulated as~\cite{2012MakhfudzP217201217201,2015BuettnerP225228}
\begin{equation}
	\mathcal{M}\ddot{\boldsymbol{\mathcal{R}}}_c+\boldsymbol{v}\times\boldsymbol{G}+k\boldsymbol{\mathcal{R}}_c=0
	\label{ThieleEqKR}
\end{equation}
with $\boldsymbol{v}=\dot{\boldsymbol{\mathcal{R}}}_c$. The third term corresponds to the restoring force arising from the harmonic potential $U=1/2k(X^2+Y^2)$.

\subsection{\label{TCDGM}The classical dynamics of the gyration mode}
\begin{table*}[t]
	\caption{\label{SMGMPara}
		The table lists the parameters used in Fig.~\ref{PRRFig1}. Apart from the parameters listed in the table, the other parameters are as follows: saturation magnetization $M_S=10^6~{\rm{A/m}}$, length of side $30~{\rm{nm}}$, disk thickness $h_G=1~{\rm{nm}}$, skyrmion chirality $\mathcal{C}=1$, skyrmion core polarity $\mathcal{P}=-1$, and skyrmion topological charge $Q=-1$.
	}
	\begin{ruledtabular}
		\begin{tabular}{lccccc}
			\textrm{No.}&
			\textrm{$A_{\rm{ex}}~({\rm{10^{-11}~J/m}})$}&
			\textrm{$D~({10^{-3}~\rm{J/m^2}})$}&
			\textrm{$K~({\rm{10^{6}~J/m^3}})$}&
			\textrm{$B_z~({\rm{mT}})$}&
			\textrm{$\alpha$}\\
			\colrule
			Figs.~\ref{PRRFig1} & 1.5 & 3 & 1 & 50 & $10^{-4}$
		\end{tabular}
	\end{ruledtabular}
\end{table*}

\begin{figure*}
	\centering
	\includegraphics[width=1\textwidth]{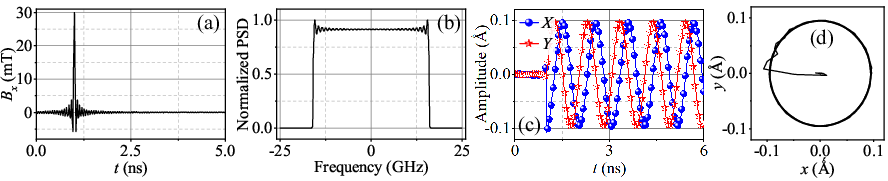}
	\caption{The waveforms of the in-plane magnetic field pulse applied are shown in both the time domain and frequency domain in (a) and (b), respectively. (c) shows the evolution of the collective coordinates $X$ and $Y$ over time. The trajectory of the clockwise gyration mode of the skyrmion is plotted in (d).}
	\label{PRRFig1}
\end{figure*}

In this work, we focus on Bloch skyrmions characterized by a chirality $\mathcal{C}=1$, core polarity $\mathcal{P}=-1$, and a resulting topological charge $Q=-1$.
The clockwise gyration mode of the skyrmion is analyzed by taking the parameters in Table~\ref{SMGMPara} and using micromagnetic simulations.
The micromagnetic simulations are performed using the OOMMF software~\cite{OOMMF} and the Ubermag Python package~\cite{2022BegP15}.
Without loss of generality, we first prepare a stable Bloch skyrmion in a square disk.
The dynamical modes of the skyrmion can be excited by magnetic fields, strain fields, and so on.
Here the excitation is performed with an in-plane magnetic field defined as $\boldsymbol{H}=(H_x(t),0,0)$, where
\begin{equation}
	H_x\left(t\right)=H_0 {\rm{sinc}}\left[2\pi f_x \left(t-t_0\right)\right].
\end{equation}
The driving amplitude, the cutoff frequency and the time shift are $B_x=\mu_0H_0=30~{\rm{mT}}$, $f_x=16~{\rm{GHz}}$, and $t_0=1~{\rm{ns}}$, respectively.
The time-domain image and the frequency-domain image of the magnetic field pulse are shown in Figs.~\ref{PRRFig1}(a) and (b), respectively.
For the micromagnetic simulation results, the evolution of the skyrmion center-of-mass coordinates $X(t)$ or $Y(t)$ over time is analyzed using the Fourier transform to obtain the spectrum of the skyrmion gyration modes.
The results of the micromagnetic simulations are displayed in Fig.~\ref{PRRFig1}.

\begin{figure*}[bt]
	\centering
	\includegraphics[width=0.85\textwidth]{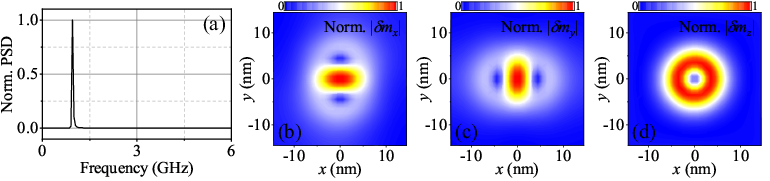}
	\caption{(a) is the spectrum of the clockwise gyration mode of the skyrmion, and the corresponding normalized mode functions of this mode are shown in (b), (c), and (d).}
	\label{PRRFig2}
\end{figure*}

Figs.~\ref{PRRFig1}(c) and (d) depict the time evolution of the skyrmion's center-of-mass coordinates and its motion trajectory within the disk, respectively.
From the spectrum in Fig.~\ref{PRRFig2}(a), the resonance frequency of the gyration mode can be extracted as $0.95~{\rm{GHz}}$.
The mode function corresponding to each eigenfrequency can be obtained by site-to-site Fourier transform technique~\cite{2016MruczkiewiczP174429174429,2020LiuP7522275222}.
The normalized mode functions of the skyrmion clockwise gyration mode are shown in Figs.~\ref{PRRFig2}(b-d).
This is consistent with our theoretical derivation in Sec.~\ref{QMFSGMApprox}.

\subsection{\label{TRFoSGM}The resonant frequency of the skyrmion gyration mode}

\begin{figure*}[bt]
	\centering
	\includegraphics[width=1\textwidth]{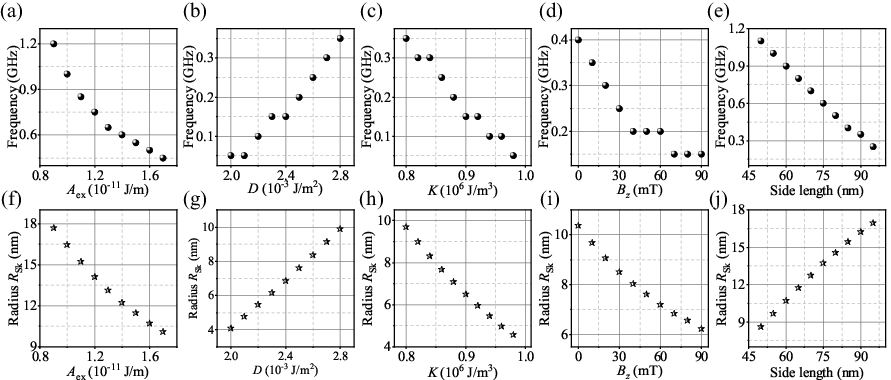}
	\caption{The dependence of the clockwise gyration mode frequency on the exchange interaction $A_{\rm{ex}}$, DM interaction $D$, uniaxial anisotropy $K$, applied magnetic field $B_z$, and disk geometry is presented. The parameters used are presented in Table~\ref{SimulationPara}.}
	\label{PRRFig3}
\end{figure*}

\begin{table*}
	\caption{\label{SimulationPara}
		The table lists the parameters used in Fig.~\ref{PRRFig3}. Apart from the varying parameters listed in the table, the other parameters are as follows: saturation magnetization $M_S=10^6~{\rm{A/m}}$, disk thickness $h_G=1~{\rm{nm}}$, skyrmion chirality $\mathcal{C}=1$, skyrmion core polarity $\mathcal{P}=-1$, and skyrmion topological charge $Q=-1$. For chiral magnetic materials, the Gilbert damping parameter is taken here as $\alpha=10^{-4}$.
	}
	\begin{ruledtabular}
		\begin{tabular}{lccccc}
			\textrm{No.}&
			\textrm{$A_{\rm{ex}}~({\rm{10^{-11}~J/m}})$}&
			\textrm{$D~({10^{-3}~\rm{J/m^2}})$}&
			\textrm{$K~({\rm{10^{6}~J/m^3}})$}&
			\textrm{$B_z~({\rm{mT}})$}&
			\textrm{Length of Side $({\rm{nm}})$}\\
			\colrule
			Figs.~\ref{PRRFig3}(a, f) & 0.9 \textendash 1.7 & 3 & 1 & 50 & 70\\
			Figs.~\ref{PRRFig3}(b, g) & 1.5 & 2 \textendash 2.8 & 1 & 50 & 70\\
			Figs.~\ref{PRRFig3}(c, h) & 1.5 & 2 & 0.8 \textendash 0.98 & 50 & 70\\
			Figs.~\ref{PRRFig3}(d, i) & 1.5 & 2.5 & 1 & 0 \textendash 90 & 70\\
			Figs.~\ref{PRRFig3}(e, j) & 1.5 & 3 & 1 & 30 & 50 \textendash 95
		\end{tabular}
	\end{ruledtabular}
\end{table*}

Next, micromagnetic simulations are employed to quantitatively examine the relationship between the gyration mode frequencies and the material parameters as well as the geometry.
Here, exchange interactions, DM interactions, $z$-direction uniaxial anisotropy, Zeeman interactions, and static magnetic interactions of the magnetic material are considered, i.e., the total energy of the system is $E_{\rm{tot}}=E_{\rm{ex}}+E_D+E_{\rm{ani}}+E_{\rm{ze}}+E_{\rm{De}}$.
First we prepare a Bloch skyrmion in a square disk and then apply an in-plane magnetic field pulse to excite the dynamical mode of the skyrmion.
The in-plane magnetic field pulse is defined as $\boldsymbol{H}=(H_x(t),0,0)$, where $H_x\left(t\right)=H_0 {\rm{sinc}}\left[2\pi f_x \left(t-t_0\right)\right]$.
The driving amplitude, the cutoff frequency and the time shift are $B_x=\mu_0H_0=5~{\rm{mT}}$, $f_x=16~{\rm{GHz}}$, and $t_0=1~{\rm{ns}}$, respectively.
All parameters used in the micromagnetic simulation are listed in Table~\ref{SimulationPara}, and all the results of the simulations are presented in Fig.~\ref{PRRFig3}.

Using micromagnetic simulations, we quantitatively demonstrate that the eigenfrequency of the skyrmion’s clockwise gyration mode is inversely correlated with the exchange interaction strength, uniaxial anisotropy, and external magnetic field, while exhibiting a positive dependence on the DM interaction~[Figs.~\ref{PRRFig3}(a-d)].
The skyrmion radius decreases with increasing exchange interaction, anisotropy, or magnetic field, but expands with enhanced DM interaction~[Figs.~\ref{PRRFig3}(f-i)].
Importantly, geometric effects introduce distinct scaling behaviors~[Figs.~\ref{PRRFig3}(e) and (j)]: under fixed material parameters, reduced boundary confinement in larger disks leads to skyrmion expansion and a corresponding decrease in gyration frequency, whereas under fixed geometric conditions, variations in material parameters result in a positive correlation between skyrmion radius and gyration frequency.
These results illustrate the significant impact of the relative scale between the skyrmion size and the system geometry on its dynamics.

\subsection{\label{DSGMMS}The dissipation of skyrmion gyroscopic modes: micromagnetic simulations}

\begin{table*}
	\caption{\label{DissipationPara}
		The table lists the parameters used in Fig.~\ref{PRRFig4}. Apart from the varying parameters listed in the table, the constant shared parameters are as follows: saturation magnetization $M_S=10^6~{\rm{A/m}}$, length of side $70~{\rm{nm}}$, disk thickness $h_G=1~{\rm{nm}}$, skyrmion chirality $\mathcal{C}=1$, skyrmion core polarity $\mathcal{P}=-1$, and skyrmion topological charge $Q=-1$.
	}
	\begin{ruledtabular}
		\begin{tabular}{lccccc}
			\textrm{No.}&
			\textrm{$A_{\rm{ex}}~({\rm{10^{-11}~J/m}})$}&
			\textrm{$D~({10^{-3}~\rm{J/m^2}})$}&
			\textrm{$K~({\rm{10^{6}~J/m^3}})$}&
			\textrm{$B_z~({\rm{mT}})$}&
			\textrm{$\alpha$}\\
			\colrule
			Fig.~\ref{PRRFig4} & 1.5 & 3 & 1 & 30 & $10^{-4}$ \textendash $10^{-1}$\\
		\end{tabular}
	\end{ruledtabular}
\end{table*}

\begin{figure}[bt]
	\centering
	\includegraphics[width=0.48\textwidth]{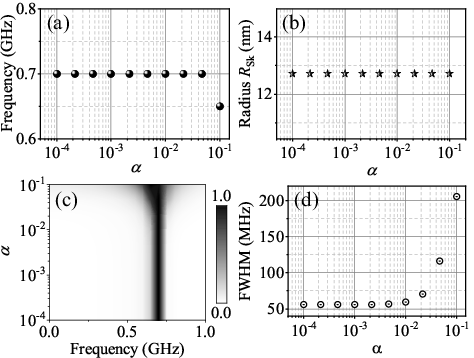}
	\caption{Panels (a) and (b) illustrate the dependence of the resonance frequency of the clockwise gyration mode and the skyrmion radius on the Gilbert damping parameter $\alpha$, respectively. The dependence of the spectrum and FWHM on the Gilbert damping parameter $\alpha$ is shown in (c) and (d). The parameters used are presented in Table~\ref{DissipationPara}.}
	\label{PRRFig4}
\end{figure}

The Thiele equation~(\ref{ThieleEqKR}) governs the skyrmion gyration mode, with the dissipation described by the dissipation tensor $\bar{\bar{D}}$.
The dissipation tensor $\bar{\bar{D}}$ is associated with the phenomenological Gilbert damping parameter $\alpha$, which can be measured experimentally.
The Gilbert damping parameter describes the dissipation mechanism arising from the coupling of the skyrmion to other degrees of freedom, such as electrons, phonons, and magnons~\cite{2017PsaroudakiP4104541045}.
To analyze the dissipation of the skyrmion gyration mode, we use the parameters listed in Table~\ref{DissipationPara} and perform micromagnetism simulations, the results of which are displayed in Fig.~\ref{PRRFig4}.

As shown in Fig.~\ref{PRRFig4}(a), when the parameters and geometry of the magnetic material are fixed, the frequency of the clockwise gyration mode of the skyrmion does not change with variations in the Gilbert damping parameter.
However, when the Gilbert damping parameter reaches $0.1$, we observe a slight decrease in the eigenfrequency of the gyration mode.
This can be attributed to the increased spectral broadening caused by the larger Gilbert damping parameter, which leads to a shift in the eigenfrequency.
In addition, Figure~\ref{PRRFig4}(b) shows that the size of the skyrmion remains unaffected by changes in the Gilbert damping parameter.

The dissipation of the gyration mode is analyzed via the spectrum, where the full width at half maximum (FWHM) of the resonance peak reflects the system's energy dissipation rate.
Specifically, a larger FWHM corresponds to faster energy loss in the system, indicating stronger dissipation.
Figure~\ref{PRRFig4}(c) presents the spectrum of the clockwise gyration mode as a function of the Gilbert damping parameter.
Notably, the bandwidth of the gyration mode increases significantly only when the Gilbert damping parameter exceeds $10^{-2}$.
As the Gilbert damping parameter increases, the FWHM also increases, with a significant rise occurring only when the parameter $\alpha$ exceeds $10^{-2}$ [Fig.~\ref{PRRFig4}(d)].

The Gilbert damping parameter of \ce{Cu_2OSeO_3} can reach $10^{-4}$ at a temperature of $5~{\rm{K}}$~\cite{2017StasinopoulosP3240832408}.
Thus, the system's dissipation can be estimated to be approximately $50~{\rm{MHz}}$.
As the temperature increases, the Gilbert damping parameter also increases, exceeding $10^{-2}$ at approximately $57~{\rm{K}}$~\cite{2017StasinopoulosP3240832408}.
The dissipation of the skyrmion gyration mode in the magnetic material \ce{Cu_2OSeO_3} has been measured to be approximately $100~{\rm{MHz}}$ ($57~\mathrm{K}$)~\cite{2021KhanP100402100402, 2021LiensbergerP100415100415}, in good agreement with the results from our simulations~[Fig.~\ref{PRRFig4}(d)].
Additionally, selecting materials without DM interactions is expected to further lower the Gilbert damping parameter, thereby reducing the system's dissipation.
Moreover, with the quantum device operating in the ${\rm{mK}}$ range, the dissipation of the gyration mode is expected to decrease further in this ultra-low temperature environment.

\section{\label{QPSGM}Quantum properties of the skyrmion gyration mode}
\subsection{\label{GMLevel}Quantized gyration modes: Harmonic oscillator-like objects}
In Sec.~\ref{CPSGM}, we examined classical gyration modes. We now turn our attention to the analysis of quantizing the gyration modes, which exhibit characteristics similar to those of quantum harmonic oscillators.
As discussed earlier, within the framework of the rigid body model, the coordinates $\boldsymbol{\mathcal{R}}_c=(X,Y)$ of the skyrmion's center satisfy the Thiele equation~(\ref{ThieleEqKR})~\cite{2017HanP,2012MakhfudzP217201217201}.
Introducing the gauge potential $\boldsymbol{A}=G(A_X,A_Y,0)$, the Hamiltonian corresponding to the Thiele equation can be expressed as
\begin{equation}
	H_{\rm{GM}}=\frac{\left(p_X+GA_X\right)^2}{2\mathcal{M}}+\frac{\left(p_Y+GA_Y\right)^2}{2\mathcal{M}}+U,
\end{equation}
where the gauge potential $\boldsymbol{A}$ satisfies the condition $\nabla\times\boldsymbol{A}=\boldsymbol{G}$, specifically fulfilling the equation $\partial_XA_Y-\partial_YA_X=1$.

To clarify the physical interpretation of the gyration mode, we select the symmetric gauge potential $\boldsymbol{A}=G(-Y,X,0)/2$ for a more detailed analysis.
In the symmetric gauge potential, the Hamiltonian of the system can be expressed as
\begin{equation}
	H_{\rm{GM}}=\frac{\left(p_X-G^\prime Y\right)^2}{2\mathcal{M}}+\frac{\left(p_Y+G^\prime X\right)^2}{2\mathcal{M}}+\frac{1}{2}k\left(X^2+Y^2\right)
	\label{HGMS}
\end{equation}
with $G^\prime=G/2$.
Expanding the Hamiltonian~(\ref{HGMS}) and utilizing the definition of orbital momentum $\boldsymbol{L}=\boldsymbol{r}\times\boldsymbol{p}$, we get
\begin{equation}
	H_{\rm{GM}}=\frac{p_X^2}{2\mathcal{M}}+\frac{p_Y^2}{2\mathcal{M}}+\frac{1}{2}\mathcal{M}\omega^2\left(X^2+Y^2\right)+\omega^\prime_0 L_z,
	\label{HGML}
\end{equation}
where $\omega^2={\omega_0^\prime}^2+\omega_h^2$, $\omega^\prime_0=G^\prime/\mathcal{M}$, and $\omega_h=\sqrt{k/\mathcal{M}}$.

Next we discuss the quantization of the Hamiltonian~(\ref{HGML}).
For convenience, in the subsequent analysis we decompose the Hamiltonian into two parts: A two-dimensional harmonic oscillator Hamiltonian and an orbital momentum Hamiltonian
\begin{subequations}
	\begin{align}
		\hat{H}_{\rm{SHO}}&=\frac{\hat{p}_X^2}{2\mathcal{M}}+\frac{\hat{p}_Y^2}{2\mathcal{M}}+\frac{1}{2}\mathcal{M}\omega^2\left(\hat{X}^2+\hat{Y}^2\right), \\
		\hat{H}_L&=\omega_0^\prime\left(\hat{X}\hat{p}_Y-\hat{Y}\hat{p}_X\right).
	\end{align}
\end{subequations}
It is worth noting that the Hamiltonian $\hat{H}_{\rm{SHO}}$ and $\hat{H}_L$ commute, $[\hat{H}_{\rm{SHO}},\hat{H}_L]=0$, implying that they share a common eigenstate.
Introducing the annihilation operators in the $x$ and $y$ directions, $\hat{a}_X=1/\sqrt{2}[\beta \hat{X}+i\hat{p}_X/(\hslash\beta)]$ and $\hat{a}_Y=1/\sqrt{2}[\beta \hat{Y}+i\hat{p}_Y/(\hslash\beta)]$, we can obtain
\begin{subequations}
	\begin{align}
		\hat{H}_{\rm{SHO}}&=\hslash\omega\left(\hat{a}_X^\dagger\hat{a}_X+\hat{a}_Y^\dagger\hat{a}_Y+1\right), \\
		\hat{H}_L&=i\hslash\omega_0^\prime\left(\hat{a}_X\hat{a}_Y^\dagger-\hat{a}_X^\dagger\hat{a}_Y\right),
	\end{align}
\end{subequations}
where $\beta\equiv\sqrt{m\omega/\hbar}$.
We define $\hat{a}_\mathrm{CW}=1/\sqrt{2}(\hat{a}_X-i\hat{a}_Y)$ and $\hat{a}_\mathrm{CCW}=1/\sqrt{2}(\hat{a}_X+i\hat{a}_Y)$, corresponding to the annihilation operators for the clockwise (CW) and counterclockwise (CCW) gyration modes, respectively.
The operators $\hat{a}_1$ and $\hat{a}_2$ satisfy the commutation relations $[\hat{a}_\mathrm{CW},\hat{a}_\mathrm{CW}^\dagger]=1=[\hat{a}_\mathrm{CCW},\hat{a}_\mathrm{CCW}^\dagger]$.
Thus, the Hamiltonian $\hat{H}_{\rm{GM}}=\hat{H}_{\rm{SHO}}+\hat{H}_L$ can be simplified to
\begin{equation}
	\hat{H}_{\rm{GM}}=\hslash\omega_{\rm{CW}}\hat{a}_\mathrm{CW}^\dagger\hat{a}_\mathrm{CW}+\hslash\omega_{\rm{CCW}}\hat{a}_\mathrm{CCW}^\dagger\hat{a}_\mathrm{CCW},
\end{equation}
where $\omega_{\rm{CW}}=\omega+\omega_0^\prime$ and $\omega_{\rm{CCW}}=\omega-\omega_0^\prime$.
Based on the analysis in Sec.~\ref{TCDGM}, we can conclude that $G\propto Q<0\rightarrow \omega_0^\prime<0$, i.e., the eigenfrequency of the CW gyration mode is smaller than that of the CCW gyration mode for Bloch skyrmions with core polarity $\mathcal{P}=-1$.
Here, we consider only the CW gyration mode $\hat{a}_\mathrm{CW}$ and, for convenience, drop the subscripts from the operators [Fig.~\ref{PRRFig1}(c, d)].
The free Hamiltonian can then be written as
\begin{equation}
	\hat{H}_{\rm{GM}}=\omega_{\rm{GM}}\hat{a}^\dagger\hat{a},
\end{equation}
where $\omega_{\rm{GM}}\equiv\omega_{\rm{CW}}$ denotes the gyration frequency.

\subsection{\label{QMFSGMApprox}Quantum magnetization fluctuations due to skyrmion gyration modes}
Next, we focus on the quantum magnetization fluctuation induced by the gyration mode.
Based on the analysis in Sec.~\ref{CPSGM}, we know that the magnetization at any moment can be expressed as $\boldsymbol{m}(\boldsymbol{r},t)$.
When the gyration mode is excited, the coordinates of the skyrmion center can be expressed as $\boldsymbol{\mathcal{R}}_c(t)=r_c[\cos (\omega_{\rm{GM}}t)\boldsymbol{e}_x+\mathcal{P}\sin (\omega_{\rm{GM}}t)\boldsymbol{e}_y]$.
Then the magnetization at time $t$ can be written as $\boldsymbol{m}(\boldsymbol{r},t)=\boldsymbol{m}[\boldsymbol{r}-\boldsymbol{\mathcal{R}}_c(t)]\approx \boldsymbol{m}(\boldsymbol{r})-[\boldsymbol{\mathcal{R}}_c(t)\cdot\nabla]\boldsymbol{m}(\boldsymbol{r})$.
The variation of the magnetization due to the excitation of the gyration mode can be expressed as
$\delta\boldsymbol{m}_l(\boldsymbol{r},t)=\delta \boldsymbol{m}_l(\boldsymbol{r})\exp(i\omega_lt)+\mathrm{c.c.}$, where $\delta\boldsymbol{m}_l(\boldsymbol{r})$ represents the mode function of the gyration mode~\cite{2020LiuP7522275222,2016MruczkiewiczP174429174429}.
This equation holds under the condition $\vert \delta \boldsymbol{m}_l(\boldsymbol{r})\vert \ll 1$.
Therefore, the magnetization field can be regarded as a superposition of a series of eigenmodes, represented as $\delta\boldsymbol{m}(\boldsymbol{r},t)=\sum_{l=\mathrm{CW},\mathrm{CCW}}\delta \boldsymbol{m}_l(\boldsymbol{r})\exp(i\omega_lt)+\mathrm{c.c.}$.
Here, we focus on the CW mode $\hat{a}$, with the subscripts omitted for simplicity
\begin{equation}
	\delta\boldsymbol{m}\left(\boldsymbol{r},t\right)=\delta\boldsymbol{m}\left(\boldsymbol{r}\right)e^{i\omega_{\rm{GM}}t}+\mathrm{c.c.},
\end{equation}
where $	\delta\boldsymbol{m}\left(\boldsymbol{r}\right)=(-r_c/2)[\partial_x\boldsymbol{m}(\boldsymbol{r})-i\mathcal{P}\partial_y\boldsymbol{m}(\boldsymbol{r})]$.
Quantizing $\delta\boldsymbol{m}(\boldsymbol{r},t)$ yields a quantized magnetization intensity
\begin{equation}
	\hat{\boldsymbol{m}}\equiv\delta \boldsymbol{m}\left(\boldsymbol{r}\right)\hat{a}+\delta \boldsymbol{m}^*\left(\boldsymbol{r}\right)\hat{a}^\dagger.
	\label{QDeltaM}
\end{equation}

\begin{figure*}[bt]
	\centering
	\includegraphics[width=0.7\textwidth]{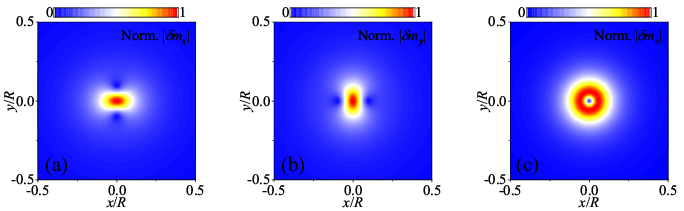}
	\caption{(a), (b), and (c) show the components of the normalized mode function $\vert\delta m_x\vert$, $\vert \delta m_y\vert$ and $\vert \delta m_z\vert$ of the skyrmion described by Eq.~(\ref{DeltaM}), respectively.}
	\label{PRRFig5}
\end{figure*}

The amplitude of the gyration mode, denoted as $r_c$, will be determined in the subsequent analysis.
Utilizing the BP model of the skyrmion Eq.~(\ref{BPmodel}), a specific expression for the modal function $\delta\boldsymbol{m}=(\delta m_x,\delta m_y, \delta m_z)$ can be computed, which is given as follows:
\begin{subequations}
	\begin{align}
		\delta m_x &= -\frac{r_c}{2}\left\{-\frac{2iR_\mathrm{Sk}\left[R_\mathrm{Sk}^2+\left(x+iy\right)^2\right]}{\left[R_\mathrm{Sk}^2+\left(x^2+y^2\right)\right]^2}\right\}, \\
		\delta m_y &= -\frac{r_c}{2}\left\{\frac{2R_\mathrm{Sk}\left[R_\mathrm{Sk}^2-\left(x+iy\right)^2\right]}{\left[R_\mathrm{Sk}^2+\left(x^2+y^2\right)\right]^2}\right\}, \\
		\delta m_z &= -\frac{r_c}{2}\left\{\frac{4R_\mathrm{Sk}^2\left(x+iy\right)}{\left[R_\mathrm{Sk}^2+\left(x^2+y^2\right)\right]^2}\right\}.
	\end{align}
	\label{DeltaM}
\end{subequations}
Figure~\ref{PRRFig5} illustrates the distribution of the mode function described by Eq.~(\ref{DeltaM}), which is consistent with the micromagnetic simulations (Fig.~\ref{PRRFig2}).

Next, we identify the gyration radius $r_c$ corresponding to a single quantum excitation of the gyration mode.
The local magnetization can be written as~\cite{2017HanP,2018GrafP241406241406,2022WeberP10251030}
\begin{equation}
	\begin{split}
		\boldsymbol{m}_{\rm{ex}}(\boldsymbol{r})&=\boldsymbol{e}_3\sqrt{1-\frac{2g\mu_B}{M_S}\vert \psi\left(\boldsymbol{r}\right)\vert^2}\\ &+\sqrt{\frac{g\mu_B}{M_S}}\left[\psi\left(\boldsymbol{r}\right)\boldsymbol{e}_++\psi^*\left(\boldsymbol{r}\right)\boldsymbol{e}_-\right],
	\end{split}
	\label{Mpsi}
\end{equation}
where $(\boldsymbol{e}_1,\boldsymbol{e}_2,\boldsymbol{e}_3)$ represents the local orthogonal basis, and $\boldsymbol{e}_3=\boldsymbol{m}(\boldsymbol{r})$ corresponds to the local equilibrium direction.
The vectors $\boldsymbol{e}_\pm$ are defined as $\boldsymbol{e}_\pm=(\boldsymbol{e}_1\pm i\boldsymbol{e}_2)/\sqrt{2}$.
The complex amplitude of the spin wave is represented by $\psi(\boldsymbol{r})$, and the probability density of magnon excitations is given by $\vert \psi(\boldsymbol{r})\vert^2$.
To determine the local orthogonal basis $(\boldsymbol{e}_1,\boldsymbol{e}_2,\boldsymbol{e}_3)$, we first simplify the expression $\boldsymbol{e}_3=\boldsymbol{m}(\boldsymbol{r})=\cos\Phi \boldsymbol{e}_\phi + \sin\Phi\boldsymbol{e}_z$, where $\boldsymbol{e}_3$ is defined in the plane spanned by $\boldsymbol{e}_\phi$ and $\boldsymbol{e}_z$.
Then we have $\boldsymbol{e}_2=\sin \Phi\boldsymbol{e}_\phi-\cos\Phi\boldsymbol{e}_z$, and the third local basis vector is given by $\boldsymbol{e}_1=\boldsymbol{e}_2\times\boldsymbol{e}_3=\boldsymbol{e}_\rho$.
From Eq.~(\ref{Mpsi}), we obtain $\delta \boldsymbol{m}(\boldsymbol{r})=\sqrt{g\mu_B/M_S}[\psi(\boldsymbol{r})\boldsymbol{e}_++\psi^*(\boldsymbol{r})\boldsymbol{e}_-]$, from which we determine the wave function $\psi\left(\boldsymbol{r}\right)$ in terms of the mode function
\begin{equation}
	\begin{split}
		\psi\left(\boldsymbol{r}\right)&=\sqrt{\frac{M_S}{g\mu_B}}\frac{\delta\boldsymbol{m}\left(\boldsymbol{r}\right)}{2}\cdot\boldsymbol{e}_+\\
		&=-\frac{r_c R_\mathrm{Sk}}{\sqrt{2}}\sqrt{\frac{M_S}{g\mu_B}}\frac{-i\cos\phi+\sin\phi}{R_\mathrm{Sk}^2+x^2+y^2}
	\end{split}
	\label{PsiDeltaM}
\end{equation}
The wave function satisfies the normalization condition over the entire disk, namely, $\int d\boldsymbol{r} \vert\psi(\boldsymbol{r})\vert^2=1$.
Thus, the gyration radius corresponding to a single excitation of the gyration mode is given by
\begin{equation}
	r_c = \sqrt{\frac{2g\mu_B\left(R^2+R_\mathrm{Sk}^2\right)}{\pi h_G R^2 M_S}}.
\end{equation}
Using the parameters $R_\mathrm{Sk}/R=0.1$, $h_G=5~{\rm{nm}}$, and $M_S=10^6~{\rm{A/m}}$, we get $r_c\approx 0.5~{\rm{\AA}}$.

\subsection{\label{MFDisk}Quantized magnetic fields generated by the excitation of the gyration mode}
In Sec.~\ref{QMFSGMApprox}, we derived that the quantum magnetization fluctuation resulting from gyration mode excitation is $\hat{\boldsymbol{m}}\equiv\delta \boldsymbol{m}\left(\boldsymbol{r}\right)\hat{a}+\delta \boldsymbol{m}^*\left(\boldsymbol{r}\right)\hat{a}^\dagger$.
The quantized magnetic field generated by the magnetization $\hat{\boldsymbol{m}}$ is
\begin{equation}
	\boldsymbol{\mathcal{B}}\left(\boldsymbol{r}\right)=\mu_0M_s\int d\boldsymbol{r}^\prime \Gamma(\boldsymbol{r},\boldsymbol{r}^\prime)\hat{\boldsymbol{m}}\left(\boldsymbol{r}^\prime\right),
	\label{Bformula}
\end{equation}
with the tensorial magnetostatic Green function $\Gamma(\boldsymbol{r},\boldsymbol{r}^\prime)=-\nabla \nabla^\prime G(\boldsymbol{r},\boldsymbol{r}^\prime)$.
The operators $\nabla$ and $\nabla^\prime$ are defined as $\nabla=(\partial_x,\partial_y,\partial_z)$ and $\nabla^\prime=(\partial_{x^\prime},\partial_{y^\prime},\partial_{z^\prime})$, respectively, and $G(\boldsymbol{r},\boldsymbol{r}^\prime)=1/(4\pi\vert \boldsymbol{r}-\boldsymbol{r}^\prime\vert)$ represents the Coulombic Green’s function.
Defining the mode function of the quantized magnetic field as
\begin{equation}
	\tilde{\boldsymbol{\mathcal{B}}}=\mu_0 M_S \int d\boldsymbol{r}^\prime \Gamma(\boldsymbol{r},\boldsymbol{r}^\prime) \delta\boldsymbol{m}\left(\boldsymbol{r^\prime}\right),
\end{equation}
Eq.~(\ref{Bformula}) can then be simplified to
\begin{equation}
	\hat{\boldsymbol{\mathcal{B}}}=\tilde{\boldsymbol{\mathcal{B}}}\hat{a}+\tilde{\boldsymbol{\mathcal{B}}}^*\hat{a}^\dagger.
	\label{MFQuantum}
\end{equation}
For the general case of $\delta \boldsymbol{m}$, the mode function can be simplified to
\begin{equation}
	\tilde{\boldsymbol{\mathcal{B}}}=\frac{\mu_0 M_S}{4\pi}\int d\boldsymbol{r}^\prime \left\{\frac{3(\boldsymbol{r}-\boldsymbol{r}^\prime)\left[\delta \boldsymbol{m}\cdot(\boldsymbol{r}-\boldsymbol{r}^\prime)\right]}{\vert \boldsymbol{r}-\boldsymbol{r}^\prime\vert^5}-\frac{\delta \boldsymbol{m}}{\vert \boldsymbol{r}-\boldsymbol{r}^\prime\vert^3}\right\}.
\end{equation}
The distance between the field point $\boldsymbol{r}=(x,y,z)$ and the source point $\boldsymbol{r}^\prime=(x^\prime,y^\prime,z^\prime)$ is given by $r=\sqrt{(x-x^\prime)^2+(y-y^\prime)^2+(z-z^\prime)^2}$.
The physical interpretation of this integral is as follows: by subdividing a magnetic system into infinitesimal volume elements, treating each as a magnetic dipole, and integrating the dipole field from each element, we can derive the total magnetic field produced by the system.

\section{\label{SNHQS}Coupling between skyrmions and NV centers}
\begin{figure}
	\centering
	\includegraphics[width=0.48\textwidth]{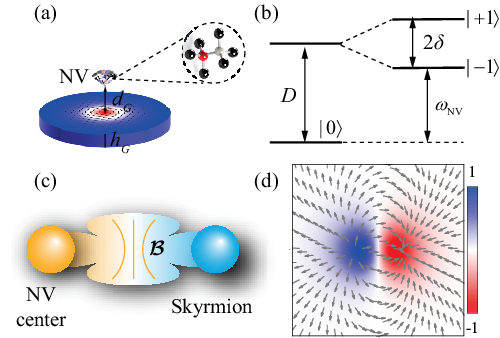}
	\caption{(a) The coupling model between the skyrmion and the NV center. (b) The energy levels of the NV center. (c) The coupling mechanism in this hybrid quantum system.
	(d) shows the distribution of the field ${\rm{Re}}(\tilde{\boldsymbol{\mathcal{B}}})$ at a position $5~{\rm{nm}}$ above a disk with a radius of $R=50~{\rm{nm}}$ and a thickness of $h_G=5~{\rm{nm}}$. Here, the reduced skyrmion radius is defined as $c=R_{\rm{Sk}}/R=0.1$, where $R_{\rm{Sk}}$ denotes the skyrmion radius.}
	\label{PRRFig6}
\end{figure}
\subsection{\label{TDE}The device}
As depicted in Fig.~\ref{PRRFig6}(a), we examine a thin disk with a radius $R$ significantly greater than its thickness $h_G$, in which the skyrmion is stabilized within the disk.
As discussed previously, gyration modes of the skyrmion exist, where the skyrmion oscillates in a circular motion around the equilibrium position, analogous to a rigid particle.
Additionally, the skyrmion gyration modes can be quantized as bosons.
A diamond particle is positioned atop the disk, with an NV center embedded within the particle.
The distance between the NV center and the disk surface is denoted by $d_G$.
The energy level structure of the NV center features a triplet state with $S=1$ [Fig.~\ref{PRRFig6}(b)].
In the presence of a magnetic field $B_z$ along the $z$ direction, the Hamiltonian of the NV center is given by $\hat{H}_{\rm{NV}}=D \hat{S}_z^2+\delta \hat{S}_z$, where $D$ represents the zero-field splitting between the subenergy levels $\vert 0\rangle$ and $\vert \pm 1\rangle$, and $\delta=\gamma_e B_z$ denotes the Zeeman splitting due to the applied magnetic field $B_z$.
Selecting the energy levels $\vert -1\rangle$ and $\vert 0\rangle$ as a qubit, the Hamiltonian of the NV center simplifies to $\hat{H}_{\rm{NV}}=\omega_{\rm{NV}}\hat{\sigma}_z/2$, with the resonant frequency $\omega_{\rm{NV}}=D-\delta$.
The Pauli operator of the NV center is defined as $\hat{\sigma}_z=\vert -1\rangle\langle -1\vert-\vert 0\rangle\langle 0\vert$.

\subsection{\label{TMFETNC}The magnetic field experienced by the NV center}
The coupling between the NV centers and the skyrmion gyration modes is achieved through the stray field around the microdisks, as shown in Fig.~\ref{PRRFig6}(c).
Fig.~\ref{PRRFig6}(d) illustrates the spatial distribution of the magnetic field mode function ${\rm{Re}}[\tilde{\boldsymbol{\mathcal{B}}}]$ above the disk.
The arrows indicate the direction of the in-plane normalized magnetic field.
The contour represents the $z$-component of the field, with positive and negative signs corresponding to the positive and negative directions along the $z$ axis.
Near the periphery of the disk, the magnetic field strength gradually decreases, and a centrosymmetric pattern in the $z$-component of the field is observed.

Considering an NV center placed directly above the disk, at the field point $\boldsymbol{r}=(0,0,H_{\rm{SN}})$ with  $H_{\rm{SN}}=d_G+h_G/2$, the magnetic field at this location can be simplified to
\begin{subequations}
	\begin{align}
		\tilde{\mathcal{B}}_x&=\frac{\mu_0 M_S}{4\pi}\int d\boldsymbol{r} \left\{\frac{3\mathcal{G}_{\rm{SN}}x}{r^5}-\frac{\delta m_x}{r^3}\right\},
		\label{Bxcenter} \\
		\tilde{\mathcal{B}}_y&=\frac{\mu_0 M_S}{4\pi}\int d\boldsymbol{r} \left\{\frac{3\mathcal{G}_{\rm{SN}}y}{r^5}-\frac{\delta m_y}{r^3}\right\},
		\label{Bycenter}\\
		\tilde{\mathcal{B}}_z&=\frac{\mu_0 M_S}{4\pi}\int d\boldsymbol{r} \left\{\frac{3\mathcal{G}_{\rm{SN}}(z-H_{\rm{SN}})}{r^5}-\frac{\delta m_z}{r^3}\right\},
		\label{Bzcenter}
	\end{align}
\end{subequations}
where $\mathcal{G}_{\rm{SN}}\equiv x\delta m_x+y\delta m_y+(z-H_{\rm{SN}})\delta m_z$. To simplify the expressions, the prime notation for the position coordinates has been excluded.

\subsection{The interaction Hamiltonian}
\begin{figure}
	\centering
	\includegraphics[width=0.45\textwidth]{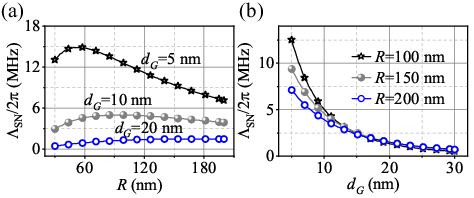}
	\caption{(a) presents the variation of the coupling strength with the disk radius $R$ while keeping $d_G$ constant, and (b) examines the variation with the distance $d_G$ for fixed radius $R$.}
	\label{PRRFig7}
\end{figure}

The interaction between the gyration mode and the NV center is mediated by magnetic dipole coupling, with the corresponding interaction Hamiltonian given by
\begin{equation}
	\hat{H}_{\rm{SN}}=-\gamma_e\hat{\boldsymbol{\mathcal{B}}}\cdot \hat{\boldsymbol{S}}=\sum_{j=x,y,z}(\lambda_j\hat{a}+\text{H.c})\hat{S}_j,
\end{equation}
where $\lambda_j=-\gamma_e\tilde{\mathcal{B}}_j$, $\gamma_e$, and $\hat{\boldsymbol{S}}=(\hat{S}_x,\hat{S}_y,\hat{S}_z)$ are the coupling strength, the gyromagnetic ratio, and the spin operator of the NV center, respectively.

Next, we derive the expressions for the coupling strengths.
First, we evaluate the coupling in the $x$ direction.
By substituting Eq.~(\ref{Bxcenter}) into the coupling strength $\lambda_x\equiv -\gamma_e\tilde{\mathcal{B}}_x$ and simplifying, we obtain
\begin{equation}
	\hat{H}_{\rm{SN}}^x=\left(\lambda_x\hat{a}+\lambda_x^*\hat{a}^\dagger\right)\hat{S}_x
\end{equation}
with the coupling strength
\begin{widetext}
	\begin{equation}
		\lambda_x=-\gamma_e \frac{\mu_0 M_S}{4 \pi R}\pi r_c \int dz \int \rho d\rho \left[-\frac{3}{r^5} \frac{2 c^2 \rho^2}{\left(c^2 + \rho^2\right)^2}\left(z-H_{\rm{SN}}\right)+i\frac{3}{r^5}\frac{c\rho^2}{c^2+\rho^2} - i\frac{2}{r^3}\frac{c^3}{\left(c^2+\rho^2\right)^2}\right],
		\label{GNgxDim}
	\end{equation}
\end{widetext}
where the integrals in azimuthal coordinates $\phi$ have been evaluated, terms with zero contribution have been omitted, and the integral has been nondimensionalized using the disk radius $R$.

Applying the same calculation for the $y$ direction, we obtain the interaction Hamiltonian
\begin{equation}
	\hat{H}_{\rm{SN}}^y=\left(\lambda_y\hat{a}+\lambda_y^*\hat{a}^\dagger\right)\hat{S}_y
\end{equation}
with coupling strength
\begin{widetext}
	\begin{equation}
		\lambda_y=-\gamma_e \frac{\mu_0 M_S}{4 \pi R}\pi r_c \int dz \int \rho d\rho \left[-i\frac{3}{r^5} \frac{2 c^2 \rho^2}{\left(c^2 + \rho^2\right)^2}\left(z-H_{\rm{SN}}\right)-\frac{3}{r^5}\frac{c\rho^2}{c^2+\rho^2} + \frac{2}{r^3}\frac{c^3}{\left(c^2+\rho^2\right)^2}\right].
		\label{GNgyDim}
	\end{equation}
\end{widetext}
For the $z$ direction coupling $\lambda_z\equiv -\gamma_e\tilde{\mathcal{B}}_z$, the calculation yields $\lambda_z=0$, indicating that the coupling strength in the $z$ direction is zero.
Comparing Eq.~(\ref{GNgxDim}) with Eq.~(\ref{GNgyDim}), we notice that ${\rm{Re}}[\lambda_x]={\rm{Im}}[\lambda_y]$ and ${\rm{Im}}[\lambda_x]=-{\rm{Re}}[\lambda_y]$.
We then define the coupling strength as $\Lambda_{\rm{SN}}=\vert \lambda_x\vert=\vert \lambda_x^*\vert=\vert \lambda_y\vert=\vert \lambda_y^*\vert$.
For convenience, defining the dimensionless integral
\begin{widetext}
	\begin{equation}
		\mathcal{F}_{\rm{SN}}=\left \vert \int dz \int \rho d\rho \left[-\frac{3}{r^5} \frac{2 c^2 \rho^2}{\left(c^2 + \rho^2\right)^2}\left(z-H_{\rm{SN}}\right)+i\frac{3}{r^5}\frac{c\rho^2}{c^2+\rho^2} - i\frac{2}{r^3}\frac{c^3}{\left(c^2+\rho^2\right)}\right] \right\vert,
		\label{EQFSN}
	\end{equation}
\end{widetext}
the coupling strength can be reduced to
\begin{equation}
	\Lambda_{\rm{SN}}=\frac{\gamma_e \mu_0 M_S r_c}{4R}\mathcal{F}_{\rm{SN}}.
\end{equation}
Based on the above analysis, we obtain $\lambda_x=\Lambda_{\rm{SN}}e^{i\phi_x}$ and $\lambda_y=\Lambda_{\rm{SN}}e^{i\phi_y}$, with phases $\phi_x=\arg (\lambda_x)$ and $\phi_y=\arg(\lambda_y)=\phi_x+\pi/2$.
The interaction Hamiltonian can then be written as $	\hat{H}_{\rm{SN}}=\Lambda_{\rm{SN}}(e^{i\phi_x}\hat{a}+e^{-i\phi_x}\hat{a}^\dagger)\hat{S}_x+i\Lambda_{\rm{SN}}(e^{i\phi_x}\hat{a}-e^{-i\phi_x}\hat{a}^\dagger)\hat{S}_y$.
Using $\hat{S}_x=(\hat{S}_++\hat{S}_-)/2$ and $\hat{S}_y=(\hat{S}_+ -\hat{S}_-)/(2i)$, and applying the rotating wave approximation, the interaction Hamiltonian can be simplified to $\hat{H}_{\rm{SN}}=\Lambda_{\rm{SN}}(e^{i\phi_x}\hat{a}\hat{S}_++e^{-i\phi_x}\hat{a}^\dagger\hat{S}_-)$.
Here, the constant phase $\phi_x$ can be neglected, as it does not affect the system dynamics.
By selecting $\vert -1\rangle$ and $\vert 0\rangle$ as the qubit, the Hamiltonian describing the coupling between the skyrmion and the NV center is given by
\begin{equation}
	\hat{H}_{\rm{SN}}=\Lambda_{\rm{SN}}\left(\hat{a}\hat{\sigma}_++\hat{a}^\dagger\hat{\sigma}_-\right),
	\label{HTGN}
\end{equation}
where the Pauli operators are defined as $\hat{\sigma}_+=\vert -1\rangle\langle0 \vert$ and $\hat{\sigma}_-=\vert 0\rangle\langle -1\vert$.
Note that if we treat the skrymion gyration mode as a classical field, the effect of the skrymion gyration mode on the NV center is equivalent to a classical drive $\hat{H}_{\text{dr}}=\Omega_d \left(\hat{\sigma}_++\hat{\sigma}_-\right)$, with the drive amplitude $\Omega_d$, rather than an interaction at the single-quantum level described by the Jaynes-Cummings (JC) model~\cite{1993ShoreP11951238}.

Figures~\ref{PRRFig7}(a) and (b) display the coupling strength $\Lambda_{\rm{SN}}$ as a function of the disk radius $R$, and the distance from the NV center to the disk surface $d_G$.
The coupling strength between the skyrmion and the NV center decreases with increasing distance $d_G$.
Here, we assume a typical saturation magnetization of $M_S=10^6~{\rm{A/m}}$~\cite{2016MruczkiewiczP174429174429,2020LiuP7522275222}.
The coupling strength between the gyration mode and NV center at the single-quantum level can reach several $\mathrm{MHz}$, potentially exceeding a dozen $\mathrm{MHz}$.

\section{\label{IBST}Coupling between skyrmions and transmon qubits.}
\begin{figure}
	\centering
	\includegraphics[width=0.48\textwidth]{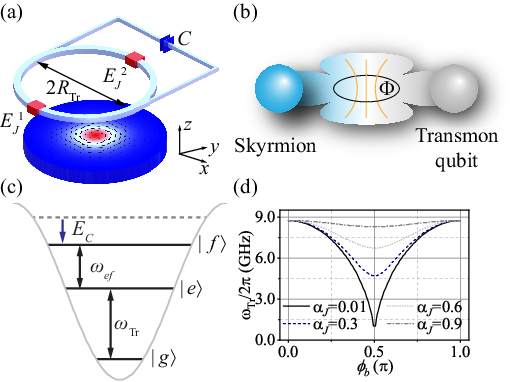}
	\caption{(a) illustrates the setup consisting of a skyrmion and a transmon qubit. Panel (b) illustrates the coupling mechanism between the skyrmions and the transmon qubits. (c) depicts the energy levels of transmon qubits. (d) plots the resonance frequencies of the transmon qubit as a function of $\alpha_J$ and $\phi_b$. The typical transmon parameters are a Josephson energy $E_J^{\rm{max}}/h=50~{\rm{GHz}}$ and a charging energy of $E_C/h=200~{\rm{MHz}}$.}
	\label{PRRFig8}
\end{figure}

\subsection{\label{VCCoMF}Calculation of the magnetic flux}
As shown in Fig.~\ref{PRRFig8}(a), we consider the configuration in which the transmon qubit is parallel to the disk.
The transmon qubit contains two Josephson junctions, $E_J^1$ and $E_J^2$, which constitute a SQUID shunted by a capacitor $C$.
By harnessing the magnetic flux generated by the field around the microdisk through the SQUID, the coupling between the skyrmion gyration mode and the transmon qubit is established, as depicted in Fig.~\ref{PRRFig8}(b).
We first calculate the magnetic flux through the SQUID, induced by the magnetic field generated by the disk.
For a SQUID placed horizontally, only the magnetic field in the $z$-direction generates an effective magnetic flux.
We define the coordinates of any point on the disk as $(x,y,z)$ and the coordinates of any point on the SQUID as $(x_f,y_f,z_f)$.
According to Eq.~(\ref{MFQuantum}), the distribution of the magnetic field in the $z$ direction on the surface of the SQUID can be simplified as
\begin{equation}
	\hat{\mathcal{B}}_z=\tilde{\mathcal{B}}_z\hat{a}+\tilde{\mathcal{B}}^*_z\hat{a}^\dagger=\frac{\mu_0 M_S}{4\pi}\left(\bar{\mathcal{B}}_z\hat{a}+\bar{\mathcal{B}}^*_z\hat{a}^\dagger\right),
\end{equation}
where $\tilde{\mathcal{B}}_z\equiv \mu_0 M_S / (4\pi)\bar{\mathcal{B}}_z$, and
\begin{equation}
	\bar{\mathcal{B}}_z=\int d\boldsymbol{r} \left\{\frac{3(z-z_f)\mathcal{G}_\mathrm{ST}}{r^5}-\frac{\delta m_z}{r^3}\right\}
	\label{BxBar}
\end{equation}
with $\mathcal{G}_\mathrm{ST}=\left[(x-x_f)\delta m_x+(y-y_f)\delta m_y+(z-z_f)\delta m_z\right]$.
The distance between the field point and the source point is denoted as $r=\sqrt{(x-x_f)^2+(y-y_f)^2+(z-z_f)^2}$.
Here we have used $\tilde{\boldsymbol{\mathcal{B}}}=[\mu_0M_S/(4\pi)]\bar{\boldsymbol{\mathcal{B}}}$.
The magnetic flux through the SQUID is given by $\hat{\Phi}=\int  \hat{\mathcal{B}}_z \boldsymbol{e}_z\cdot d\boldsymbol{S}$.
Defining $\bar{\mathcal{F}}_\Phi=\int dx_f \int d y_f \bar{\mathcal{B}}_z(x_f,y_f)=\mathcal{F}_\Phi e^{i\phi_{\rm{GM}}}$, we can obtain
\begin{equation}
	\hat{\Phi}=\frac{\mu_0 M_SR^2}{4\pi}\mathcal{F}_\Phi\left(e^{i\phi_{\rm{GM}}}\hat{a}+e^{-i\phi_{\rm{GM}}}\hat{a}^\dagger\right),
	\label{PhiQ2}
\end{equation}
where the modulus $\mathcal{F}_\Phi=\vert \bar{\mathcal{F}}_\Phi\vert$ and the phase $\phi_{\rm{GM}}=\arg (\bar{\mathcal{F}}_\Phi)$.
The constant phase $\phi_{\rm{GM}}$ typically does not influence the dynamics of the system and can thus usually be ignored.

\subsection{\label{TFHTTQ}The free Hamiltonian of the transmon qubit}
The Hamiltonian of a Transmon qubit can be written as
\begin{equation}
	\hat{H}_{\rm{TTQ}}=4E_C\hat{N}^2+\mathcal{E}_{\rm{JJ}},
	\label{HTSQ}
\end{equation}
where the inductive energy $\mathcal{E}_{\rm{JJ}}$ is defined as
\begin{equation}
	\mathcal{E}_{\rm{JJ}}=-E_J^{\rm{max}}\mathcal{S}\left(\phi_{\rm{ext}}\right)\cos\left[\hat{\delta}-\arctan\left(\alpha_J\tan\phi_{\rm{ext}}\right)\right],
	\label{EInd}
\end{equation}
which is nonlinearly related to the superconducting phase difference $\hat{\delta}$ and the magnetic flux $\phi_{\rm{ext}}=\pi \Phi_{\rm{ext}}/\Phi_0$.
The parameters involved in Eq.~(\ref{HTSQ}) and Eq.~(\ref{EInd}) are: the imbalance between the Josephson energies or SQUID asymmetry $\alpha_J=\vert E_J^1-E_J^2\vert/E_J^{\rm{max}}$ with $E_J^{\rm{max}}=E_J^1+E_J^2$; the flux quantum $\Phi_0=h/2e$ with Planck constant $h$ and elementary charge $e$; charging energy $E_C$; $\mathcal{S}(\phi_{\rm{ext}})=\sqrt{\cos^2\phi_{\rm{ext}}+\alpha_J^2\sin^2\phi_{\rm{ext}}}$; and the charge number operator $\hat{N}$ conjugate to $\hat{\delta}$.
Utilizing the trigonometric relationship
\begin{subequations}
	\begin{align}
		&\cos\left(A-B\right)=\cos\left(A\right)\cos\left(B\right)+\sin\left(A\right)\sin\left(B\right), \\
		&\cos\left(\arctan A\right)=\frac{1}{\sqrt{1+A^2}}, \\
		&\sin\left(\arctan A\right)=\frac{A}{\sqrt{1+A^2}},
	\end{align}
	\label{COSSIN}
\end{subequations}
the Hamiltonian~(\ref{HTSQ}) can be simplified to
\begin{equation}
	\hat{H}_{\rm{TTQ}}=4E_C\hat{N}^2-E_J^{\rm{max}}\vert \cos\phi_{\rm{ext}}\vert\left(\cos\hat{\delta}+\alpha_J\tan\phi_{\rm{ext}}\sin\hat{\delta}\right),
	\label{HTSQ2}
\end{equation}
where the external magnetic flux $\phi_{\rm{ext}}$ is defined as $\phi_{\rm{ext}}=\phi_b+\phi$.
The first term, $\phi_b$, denotes the extra flux to modulate the transmon qubits, and the second term, $\phi=\pi \Phi/\Phi_0$, denotes the quantized flux resulting from the excitation of the gyration mode.
With $\phi\ll \phi_b$, the Hamiltonian~(\ref{HTSQ2}) can be reduced to the free and interaction terms
\begin{subequations}
	\begin{align}
		\hat{H}_{\rm{Tr}}&=4E_C\hat{N}^2-sE_J^{\rm{max}}\left(\cos\phi_b\cos\hat{\delta}+\alpha_J\sin\phi_b\sin\hat{\delta}\right), \label{FreeTerm}\\
		\hat{H}_{\rm{ST}}&=sE_J^{\rm{max}}\phi\left(\sin\phi_b\cos\hat{\delta}-\alpha_J\cos\phi_b\sin\hat{\delta}\right), \label{GTTerm}
	\end{align}
\end{subequations}
where $s={\rm{sgn}}[\cos \phi_{\rm{ext}}]\approx{\rm{sgn}}[\cos \phi_b]$.

Next, we quantize the free Hamiltonian $\hat{H}_{\rm{Tr}}$, while the quantization of the interaction Hamiltonian $\hat{H}_{\rm{ST}}$ is discussed in Sec.~\ref{SecHGT}.
Defining $\tan x=\alpha_J \sin\phi_b/\cos\phi_b=\alpha_J\tan\phi_b$, the Hamiltonian $\hat{H}_{\rm{Tr}}$ reduces to
\begin{equation}
	\hat{H}_{\rm{Tr}}=4E_C\hat{N}^2-E_J^{\rm{max}}\mathcal{S}\left(\phi_b\right)\cos\hat{\widetilde{\delta}}
\end{equation}
with $\mathcal{S}(\phi_b)=\sqrt{\cos^2\phi_b+\alpha_J^2\sin^2\phi_b}$ and $\hat{\widetilde{\delta}}=\hat{\delta}-\arctan(\alpha_J\tan\phi_b)$.
When the qubit is in the transmon regime, i.e., $E_J^{\rm{max}}\mathcal{S}(\phi_b)\gg E_C$, and the zero-point fluctuation of the phase satisfies $\widetilde{\delta}_{\rm{zpf}}=[2E_C/(E_J^{\rm{max}}\mathcal{S}(\phi_b))]^{1/4}\ll 1$, then the Hamiltonian of the qubit can be simplified as
\begin{equation}
	\hat{H}_{\rm{Tr}}=4E_C\hat{N}^2-E_J^{\rm{max}}\mathcal{S}\left(\phi_b\right)\left(-\frac{1}{2}\hat{\widetilde{\delta}}^2+\frac{1}{24}\hat{\widetilde{\delta}}^4\right),
\end{equation}
where the constant term has been neglected.
Introducing the bosonic field operator of the qubit excitation
\begin{subequations}
	\begin{align}
		\hat{N}&=i\left[\frac{E_J^{\rm{max}}\mathcal{S}\left(\phi_b\right)}{32E_C}\right]^{\frac{1}{4}}\left(\hat{b}^\dagger-\hat{b}\right),\\
		\hat{\widetilde{\delta}}&=\left[\frac{2E_C}{E_J^{\rm{max}}\mathcal{S}\left(\phi_b\right)}\right]^{\frac{1}{4}}\left(\hat{b}^\dagger+\hat{b}\right),
	\end{align}
	\label{QNDelta}
\end{subequations}
the Hamiltonian $\hat{H}_{\rm{Tr}}$ can be written as
\begin{equation}
	\hat{H}_{\rm{Tr}}=\omega_{\rm{Tr}}\hat{b}^\dagger\hat{b}-\frac{1}{2}E_C\hat{b}^\dagger\hat{b}^\dagger\hat{b}\hat{b}.
	\label{HTrSQ}
\end{equation}
The energy levels of the Hamiltonian~(\ref{HTrSQ}) are shown in Fig.~\ref{PRRFig8}(c).
Selecting the ground state $\vert g\rangle$ and the first excited state $\vert e\rangle$ as a qubit, the Hamiltonian~(\ref{HTrSQ}) can be further simplified as
\begin{equation}
	\hat{H}_{\rm{Tr}}=\frac{\omega_{\rm{Tr}}}{2}\hat{\sigma}_z^T
	\label{HTrSQ2}
\end{equation}
with $\omega_{\rm{Tr}}=\sqrt{E_J^{\rm{max}}\mathcal{S}(\phi_b)E_C}-E_C$ and the Pauli operator $\hat{\sigma}_z^T=\vert e\rangle\langle e\vert-\vert g\rangle\langle g\vert$.
As shown in Fig.~\ref{PRRFig8}(d), the resonance frequency $\omega_{\rm{Tr}}$ can be controlled by the parameters $\alpha_J$ and $\phi_b$, allowing modulation in the GHz range.

\subsection{\label{SecHGT}The interaction Hamiltonian}
\begin{figure}
	\centering
	\includegraphics[width=0.48\textwidth]{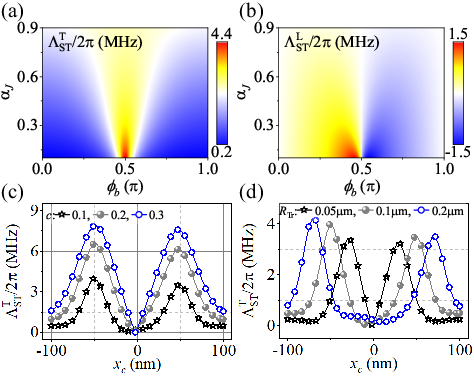}
	\caption{(a) and (b) show the variations of transverse $\Lambda_{\rm{ST}}^T$ and longitudinal $\Lambda_{\rm{ST}}^L$ couplings with the parameters of the superconducting qubit, respectively. (c) and (d) illustrate the transverse coupling strength $\Lambda_{\rm{ST}}^T$ as a function of the SQUID's spatial position, the reduced skyrmion radius $c$, and the SQUID size $R_{\rm{Tr}}$. The parameters of the disk are fixed as follows: radius $R=100~{\rm{nm}}$, thickness $h_G=5~{\rm{nm}}$, and saturation magnetization $M_S=10^6~{\rm{A/m}}$. The other parameters used here are presented in detail in Table~\ref{PRRFig9Para}. }
	\label{PRRFig9}
\end{figure}

In this section, we will discuss the quantization of the interaction Hamiltonian $\hat{H}_{\rm{ST}}$.
Using Eq.~(\ref{COSSIN}) and $\hat{\widetilde{\delta}}=\hat{\delta}-\arctan(\alpha_J\tan\phi_b)$, the interaction Hamiltonian can be simplified to
\begin{equation}
	\hat{H}_{\rm{ST}}=\frac{E_J^{\rm{max}}\phi}{\mathcal{S}\left(\phi_b\right)}\left[\frac{\sin\left(2\phi_b\right)}{2}\left(1-\alpha_J^2\right)\cos\hat{\widetilde{\delta}}-\alpha_J\sin\hat{\widetilde{\delta}}\right].
\end{equation}
With $E_J^{\rm{max}}\mathcal{S}(\phi_b)\gg E_C$, and the zero-point fluctuation of the phase $\widetilde{\delta}_{\rm{zpf}}=[2E_C/(E_J^{\rm{max}}\mathcal{S}(\phi_b))]^{1/4}\ll 1$, then the Hamiltonian $\hat{H}_{\rm{ST}}$ can be simplified as
\begin{equation}
	\hat{H}_{\rm{ST}}=\frac{E_J^{\rm{max}}\phi}{\mathcal{S}\left(\phi_b\right)}\left[\frac{\sin\left(2\phi_b\right)}{2}\left(1-\alpha_J^2\right)\left(-\frac{1}{2}\hat{\widetilde{\delta}}^2\right)-\alpha_J\hat{\widetilde{\delta}}\right],
	\label{HGT2}
\end{equation}
where the constant terms and higher order terms ($\hat{\widetilde{\delta}}^4$ and $\hat{\widetilde{\delta}}^3$) have been ignored due to their negligible effects (Appendix~\ref{EHOT}).
The transmon qubit has an anharmonic energy level structure $\{\vert g\rangle,\vert e\rangle,\vert f\rangle, \dots\}$, and we typically select the ground state $\vert g\rangle$ and the first excited state $\vert e\rangle$ to construct a qubit~(see Sec.~\ref{TFHTTQ}).
In the Hilbert space spanned by $\vert g\rangle$ and $\vert e\rangle$, the interaction Hamiltonian is quantized as

\begin{equation}
	\hat{H}_{\rm{ST}}=-\Lambda_{\rm{ST}}^L\left(\hat{a}+\hat{a}^\dagger\right)\hat{\sigma}_+^T\hat{\sigma}_-^T-\Lambda_{\rm{ST}}^T\left(\hat{a}\hat{\sigma}_+^T+\hat{a}^\dagger\hat{\sigma}_-^T\right),
\end{equation}
where the coupling strengths are defined as
\begin{subequations}
	\begin{align}
		\Lambda_{\rm{ST}}^L&=\frac{\mu_0 M_S R^2\left(1-\alpha_J^2\right)\sin\left(2\phi_b\right)}{8\hslash\Phi_0}\sqrt{\frac{2E_CE_J^{\rm{max}}}{S^3(\phi_b)}}\mathcal{F}_\Phi, \\
		\Lambda_{\rm{ST}}^T&=\frac{\mu_0 M_S R^2\alpha_J}{4\hslash\Phi_0}\left[\frac{2E_CE_J^{\rm{max}}}{S^5(\phi_b)}\right]^{1/4}\mathcal{F}_\Phi.
	\end{align}
\end{subequations}
The Pauli operators is given by $\hat{\sigma}_+^{T}=(\hat{\sigma}_-^{T})^\dagger=\vert e\rangle\langle g\vert$.
Here, the constant phase​ $\phi_\mathrm{GM}$ is neglected.
Figures~\ref{PRRFig9}(a) and (b) illustrate that the coupling strength can be modulated by adjusting $\alpha_J$ and the additional flux $\phi_b$.

\begin{table*}
	\caption{\label{PRRFig9Para}
		The table lists the parameters used in Fig.~\ref{PRRFig9}.
	}
	\begin{ruledtabular}
		\begin{tabular}{lccccccccc}
			\textrm{No.}&
			\textrm{$x_c~({\rm{nm}})$}&
			\textrm{$y_c~({\rm{nm}})$}&
			\textrm{$z_c/h_G$}&
			\textrm{$c~(R_{\rm{Sk}}/R)$}&
			\textrm{$E_J^{\rm{max}}/h~({\rm{GHz}})$}&
			\textrm{$E_C/h~({\rm{MHz}})$}&
			\textrm{$R_{\rm{Tr}}~({\rm{nm}})$}&
			\textrm{$\alpha_J$}&
			\textrm{$\phi_b$}\\
			\colrule
			Figs.~\ref{PRRFig9}(a,b) & $50$ & 0 & 2 & 0.1 & 50 & 200 & 50 & 0.1 \textendash 0.9 & 0\textendash $\pi$ \\
			Fig.~\ref{PRRFig9}(c) & $-100\textendash100$ & 0 & 3 & 0.1,0.2,0.3 & 50 & 200 & 50 & 0.06 & $\pi/2$ \\
			Fig.~\ref{PRRFig9}(d) & $-100\textendash100$ & 0 & 3 & 0.1 & 50 & 200 & 50,100,200 & 0.06 & $\pi/2$ \\
			
		\end{tabular}
	\end{ruledtabular}
\end{table*}

Next, the discussions focus on the transverse coupling strength $\Lambda_{\rm{ST}}^T$.
Since the transmon qubit and the gyration mode are coupled through the magnetic flux $\hat{\Phi}$, the coupling strength can be tuned by adjusting the spatial position of the SQUID, altering the radius of the SQUID, or increasing the skyrmion radius. These modifications can enhance the magnetic field passing through the SQUID, thereby strengthening the coupling.
Here, we use the central coordinates of the SQUID, $(x_c,y_c,z_c)$, to represent its spatial position.
Assuming $y_c=0$ and $z_c=3h_G$, it can be observed from Figs.~\ref{PRRFig9}(c) and (d) that when $x_c=0$, the coupling strength is minimized. This is attributed to the fact that the $\hat{\mathcal{B}}_z$ fields in the $+z$ and $-z$ directions simultaneously pass through the SQUID, resulting in an effective magnetic flux close to zero [Fig.~\ref{PRRFig6}(d)].
When the SQUID is offset from the disk center, the $\hat{\mathcal{B}}_z$ fields in the $+z$ and $-z$ directions passing through it are no longer equal.
At an optimal position, the magnetic field passing through the SQUID will predominantly be in either the $+z$ or $-z$ direction, maximizing the magnetic flux and thus achieving the maximum coupling strength.

In Fig.~\ref{PRRFig9}(c), as the radius of the skyrmion increases, the magnetic field passing through the SQUID intensifies, leading to an increase in magnetic flux.
Consequently, the coupling strength increases with the enlargement of the skyrmion radius.
In Fig.~\ref{PRRFig9}(d), as the radius of the SQUID $R_{\rm{Tr}}$ increases, the optimal position for coupling strength shifts outward.
This adjustment aims to ensure that the magnetic field passing through the SQUID is predominantly in a single direction.
In other words, a larger SQUID can be employed, requiring its loop to cover only half of the disk.
This is feasible because the $\hat{\mathcal{B}}_z$ field is primarily concentrated at the disk center, with nearly zero field at the edges.

\begin{figure}
	\centering
	\includegraphics[width=0.48\textwidth]{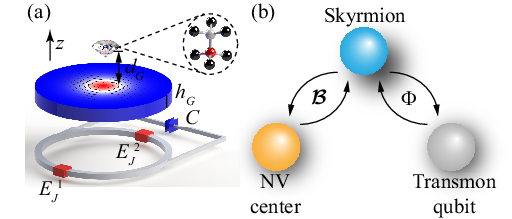}
	\caption{(a) A three-body hybrid quantum system comprising a skyrmion, an NV center, and a transmon qubit. (b) The coupling mechanism in this hybrid quantum system. }
	\label{PRRFig10}
\end{figure}

\section{\label{ICBNSQ}Indirect coupling between the NV center and the superconducting qubit}
In this section, we analyze a tripartite hybrid quantum system comprising a skyrmion, an NV center, and a transmon qubit.
As illustrated in Fig.~\ref{PRRFig10}(a), the diamond particle and the transmon qubit are positioned above and beneath the micromagnetic disk, respectively.
As discussed in Secs.~\ref{SNHQS} and \ref{IBST}, the skyrmion gyration mode is coupled to the NV center through magnetic dipole interaction, and to the transmon qubit via flux-mediated coupling.
The skyrmion gyration mode can serve as a mediator enabling both coherent and dissipative coupling between NV centers and transmon qubits.
Coherent coupling enables quantum state conversion between NV centers and transmon qubits, whereas dissipative coupling facilitates non-reciprocal quantum state conversion between them.
In the following, we provide a detailed discussion of the indirect coherent and dissipative couplings between NV centers and transmon qubits.
The longitudinal coupling term $\Lambda_{\rm{ST}}^L\left(\hat{a}+\hat{a}^\dagger\right)\hat{\sigma}_+^T\hat{\sigma}_-^T$ has been neglected due to the large detuning condition ($\omega_{\rm{GM}},\omega_{\rm{NV}},\omega_{\rm{Tr}}\gg \Lambda_{\rm{SN}},\Lambda_{\rm{ST}}^T,\Lambda_{\rm{ST}}^L$) or by modulating the qubit parameters $\{\phi_b,\alpha_J\}$.

\subsection{\label{SMCCNTMAINTEXT}Skyrmion-mediated coherent coupling between an NV center and a transmon qubit}
\begin{figure}
	\centering
	\includegraphics[width=0.45\textwidth]{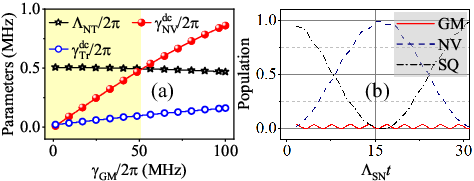}
	\caption{(a) Indirect coupling between an NV center and a transmon qubit with $\Lambda_\mathrm{SN}/2\pi=12.5~\mathrm{MHz}$ and $\Lambda_\mathrm{ST}/2\pi=5.05~\mathrm{MHz}$. (b) State conversion between NV centers and transmon qubits via the exchange of virtual excitations of the gyration mode with $\Lambda_\mathrm{SN}=\Lambda_\mathrm{ST}$. The parameters used are $\gamma_\mathrm{NV}^\mathrm{dp}/2\pi=10~\mathrm{kHz}$, $\gamma_\mathrm{NV}^\mathrm{dc}/2\pi\approx0$, and $\Delta_\mathrm{GM}=10\Lambda_\mathrm{SN}$. Here, we take the typical relaxation and dephasing times for the transmon as $T_1=T_2=50~{\rm{\mu s}}$~\cite{2020KjaergaardP369395}.}
	\label{PRRFig11}
\end{figure}

The Hamiltonian of the hybrid quantum system, comprising a skyrmion, an NV center, and a transmon qubit, is given by
\begin{equation}
	\hat{H}_\mathrm{NST}=\omega_\mathrm{GM}\hat{a}^\dagger\hat{a}+\frac{\omega_\mathrm{NV}}{2}\hat{\sigma}_z+\frac{\omega_\mathrm{Tr}}{2}\hat{\sigma}_z^T+\hat{H}_\mathrm{SN}+\hat{H}_\mathrm{ST}.
	\label{HNST}
\end{equation}
Here, $\omega_\mathrm{GM}$, $\omega_\mathrm{NV}$, and $\omega_\mathrm{Tr}$ denote the resonance frequencies of the gyration mode, NV center, and transmon qubit, respectively.
When the frequencies of the gyration modes are significantly detuned from those of the NV center and transmon qubit, their indirect coupling can be realized through the exchange of virtual excitations of the gyration modes, described by the effective Hamiltonian (see Appendix~\ref{SMCCNT})
\begin{equation}
	\hat{H}_\mathrm{NT}^\mathrm{coh}=\frac{\omega_\mathrm{NV}^\mathrm{eff}}{2}\hat{\sigma}_z+\frac{\omega_\mathrm{Tr}^\mathrm{eff}}{2}\hat{\sigma}_z^T+\Lambda_\mathrm{NT}\left(\hat{\sigma}_+\hat{\sigma}_-^T+\hat{\sigma}_+^T\hat{\sigma}_-\right),
\end{equation}
with the effective frequency $\{\omega_\mathrm{NV}^\mathrm{eff}=\omega_\mathrm{NV}-\alpha^2 \Delta_\mathrm{GM},\omega_\mathrm{Tr}^\mathrm{eff}=\omega_\mathrm{Tr}-\beta^2 \Delta_\mathrm{GM}\}$, the detuning $\Delta_\mathrm{GM}=\omega_\mathrm{GM}-\omega_\mathrm{Tr}$, and the indirect coupling strength $\Lambda_\mathrm{NT}=\Lambda_\mathrm{ST}\alpha$.
Here, the parameters $\alpha\approx\Lambda_{\rm{SN}}/\vert\Delta_\mathrm{GM}\vert$ and $\beta\approx\Lambda_{\rm{ST}}/\vert\Delta_\mathrm{GM}\vert$ satisfy $\alpha,\beta\ll 1$.
The dissipation of gyration modes increases the decay rates of NV centers and transmon qubits, with effective decay rates of $\Gamma_\mathrm{NV}^\mathrm{dc}=\gamma_\mathrm{NV}^\mathrm{dc}+\alpha^2\gamma_\mathrm{GM}$ and $\Gamma_\mathrm{Tr}^\mathrm{dc}=\gamma_\mathrm{Tr}^\mathrm{dc}+\beta^2\gamma_\mathrm{GM}$, respectively.
The symbols $\gamma_\mathrm{NV}^\mathrm{dc}$ ($\gamma_\mathrm{Tr}^\mathrm{dc}$) and $\gamma_\mathrm{NV}^\mathrm{dp}$ ($\gamma_\mathrm{Tr}^\mathrm{dp}$) represent the intrinsic decay rate and dephasing rate of the NV center (transmon qubit), respectively.
Figure~\ref{PRRFig11}(a) illustrates the dependence of the indirect coupling strength on the gyration mode dissipation.
The coupling strength satisfies $\Lambda_\mathrm{NT}>\Gamma_\mathrm{NV,Tr}^\mathrm{dc},\gamma_\mathrm{NV,Tr}^\mathrm{dp}$ in the shaded region, indicating that skyrmion-mediated indirect coupling between the NV center and the transmon qubit can reach the strong coupling regime.
State conversion between an NV center and a transmon qubit can be achieved through the exchange of virtual excitations of the gyration mode [Fig.~\ref{PRRFig11}(b)].
In other words, quantum information can be transferred from the quantum processor (transmon qubits) to the quantum memory (NV centers).

\subsection{\label{SMDCNT}Skyrmion-mediated dissipative coupling between an NV center and a transmon qubit}
We next discuss the realization of nonreciprocal interactions between NV centers and transmon qubits via skyrmion-mediated dissipative coupling.
By applying a microwave drive $\hat{H}_{\rm{qd}}=-\Omega_1(e^{i\omega_1 t}\hat{\sigma}_-+e^{-i\omega_1 t}\hat{\sigma}_+)-\Omega_2(e^{i\omega_2 t}\hat{\sigma}_-+e^{-i\omega_2 t}\hat{\sigma}_+)$ to the NV center, the Hamiltonian of the hybrid quantum system becomes
\begin{equation}
	\hat{H}_\mathrm{NSTD}=\hat{H}_\mathrm{NST}+\hat{H}_{\rm{qd}},
	\label{NSTD}
\end{equation}
where $\Omega_{1/2}$ and $\omega_{1/2}$ are the driving strengths and frequencies, respectively.
Transforming to the rotating frame of drive $\Omega_1$ yields $\hat{H}_{\rm{NSTD}}=\Delta_{\rm{NV}}/2\hat{\sigma}_z + \Delta_{\rm{Tr}}/2\hat{\sigma}_z^T + \Delta_{\rm{GM}}\hat{a}^\dagger \hat{a}+\Lambda_{\rm{SN}}(\hat{a}\hat{\sigma}_++\hat{a}^\dagger\hat{\sigma}_-)+\Lambda_{\rm{ST}}^T(\hat{a}\hat{\sigma}_+^T+\hat{a}^\dagger\hat{\sigma}_-^T)-\Omega_1(\hat{\sigma}_-+\hat{\sigma}_+)-\Omega_2[e^{i(\omega_2-\omega_1) t}\hat{\sigma}_-+e^{-i(\omega_2-\omega_1) t}\hat{\sigma}_+]$, where $\Delta_{\rm{NV}}=\omega_{\rm{NV}}-\omega_1$, $\Delta_{\rm{Tr}}=\omega_{\rm{Tr}}-\omega_1$, and $\Delta_{\rm{GM}}=\omega_{\rm{GM}}-\omega_1$.
The first drive is converted into a time-independent term, assumed to be the most significant, defined as $\hat{H}_0=-\Omega_1(\hat{\sigma}_-+\hat{\sigma}_+)$.
By employing the transformation $\hat{H}_{\rm{JR}}=\exp(i\hat{H}_0t)(\hat{H}_{\rm{NSTD}}-\hat{H}_0)\exp(-i\hat{H}_0t)$, tuning the drive frequency to satisfy $\omega_1-\omega_2=2\Omega_1$, and assuming that drive $\Omega_1$ is sufficiently strong, the Hamiltonian $\hat{H}_\mathrm{NSTD}$ can be reduced to~\cite{2012BallesterP2100721007}
\begin{equation}
	\begin{split}
		\hat{H}_{\rm{JR}}&=\frac{\Omega_2}{2}\hat{\sigma}_z+\frac{\Delta_{\rm{Tr}}}{2}\hat{\sigma_z}^T+\Delta_{\rm{GM}}\hat{a}^\dagger \hat{a} \\
		&+\bar{\Lambda}_{\rm{SN}}\left(\hat{a}+\hat{a}^\dagger\right)\hat{\sigma}_x+\Lambda_{\rm{ST}}^T\left(\hat{a}\hat{\sigma}_+^T+\hat{a}^\dagger\hat{\sigma}_-^T\right)
	\end{split}
	\label{HJR}
\end{equation}
with $\bar{\Lambda}_{\rm{SN}}=\Lambda_{\rm{SN}}/2$.
In other words, this system incorporates both JC coupling and effective Rabi coupling.

\begin{figure}[bt]
	\centering
	\includegraphics[width=0.45\textwidth]{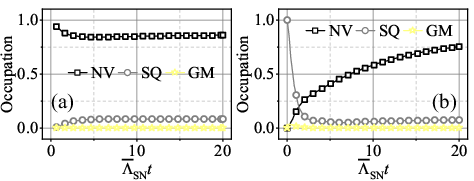}
	\caption{Nonreciprocal state conversion between NV centers and transmon qubits, with parameters $\eta_{\rm{NT}}=3$, $\Delta_\mathrm{GM}=\Omega_2=\Delta_\mathrm{Tr}=0.5\bar{\Lambda}_\mathrm{SN}$, and $\gamma_\mathrm{GM}=10\max(\bar{\Lambda}_\mathrm{SN}, \Lambda_\mathrm{ST})$.}
	\label{PRRFig12}
\end{figure}

Next, this system is utilized to achieve a nonreciprocal response between an NV center and a transmon qubit. Considering the dissipation of the gyration mode, the system's master equation is given by
\begin{equation}
	\dot{\hat{\rho}}=-i[\hat{H}_{\rm{JR}},\hat{\rho}]+\gamma_{\rm{GM}}D\left[\hat{a}\right]\hat{\rho},
	\label{MSENSTO}
\end{equation}
where the Lindblad operator is defined as $D[\hat{O}]=\hat{O}\hat{\rho}\hat{O}^\dagger-\{\hat{O}^\dagger\hat{O},\hat{\rho}\}/2$.
By applying $\partial_t\hat{O}=-i[\hat{O},\hat{H}]+1/2\{\hat{L}^\dagger[\hat{O},\hat{L}]+[\hat{L}^\dagger\hat{O}]\hat{L}\}$, the equation of motion for the operator $\hat{a}$ is obtained as
\begin{equation}
	\partial_t \hat{a}=-i\left(\Delta_{\rm{GM}}\hat{a}+\bar{\Lambda}_{\rm{SN}}\hat{\sigma}_++\bar{\Lambda}_{\rm{SN}}\hat{\sigma}_-+\Lambda_{\rm{ST}}^T\hat{\sigma}_-^T\right)-\frac{\gamma_{\rm{GM}}}{2}\hat{a}.
\end{equation}
When the dissipation of the gyration mode is large, $\gamma_{\rm{GM}}\gg\max\{\bar{\Lambda}_{\rm{SN}},\Lambda_{\rm{ST}}^T\}$, the gyration mode can be adiabatically eliminated by setting $\partial_t \hat{a}=0$. Then we can get~\cite{2022ZhangP2403824038}
\begin{equation}
	\hat{a}=\frac{-i\left(\bar{\Lambda}_{\rm{SN}}\hat{\sigma}_++\bar{\Lambda}_{\rm{SN}}\hat{\sigma}_-+\Lambda_{\rm{ST}}^T\hat{\sigma}_-^T\right)}{i\Delta_{\rm{GM}}+\gamma_{\rm{GM}}/2}.
	\label{EliminateGM}
\end{equation}
Substituting Eq.~(\ref{EliminateGM}) into the master equation Eq.~(\ref{MSENSTO}), we obtain the master equation for the dynamics involving only NV centers and transmon qubits, given by
\begin{equation}
	\dot{\hat{\rho}}=-i\left[\hat{H}_{\rm{NT}},\hat{\rho}\right]+\Gamma_{\rm{NT}}D\left[\hat{\Xi}\right]\hat{\rho}
	\label{MESNTEff}
\end{equation}
with $\hat{H}_{\rm{NT}}=\bar{\Omega}_{\rm{NV}}/2\hat{\sigma}_z+\bar{\Omega}_{\rm{Tr}}/2\hat{\sigma}_z^T-\bar{\Lambda}_{\rm{NT}}\hat{\sigma}_x\hat{\sigma}_x^T$ and $\hat{\Xi}=\hat{\sigma}_++\hat{\sigma}_-+\eta_{\rm{NT}}\hat{\sigma}_-^T$.
The effective parameters of the system are defined as $\eta_{\rm{NT}}=\Lambda_{\rm{ST}}^T/\bar{\Lambda}_{\rm{SN}}$, $\bar{\Omega}_{\rm{NV}}=\Omega_2$, $\bar{\Omega}_{\rm{Tr}}=\Delta_{\rm{Tr}}-\eta_{\rm{NT}}^2\Delta_{\rm{GM}}\bar{\Lambda}_{\rm{SN}}^2/(\Delta_{\rm{GM}}^2+\gamma_{\rm{GM}}^2/4)$, $\bar{\Lambda}_{\rm{NT}}=\eta_{\rm{NT}}\Delta_{\rm{GM}}\bar{\Lambda}_{\rm{SN}}^2/(\Delta_{\rm{GM}}^2+\gamma_{\rm{GM}}^2/4)$, and $\Gamma_{\rm{NT}}=\gamma_{\rm{GM}}\bar{\Lambda}_{\rm{SN}}^2/(\Delta_{\rm{GM}}^2+\gamma_{\rm{GM}}^2/4)$.
Subsequently, the quantum Langevin equations for the system can be written as
\begin{subequations}
	\begin{align}
		\partial_t\hat{\sigma}_-&=\left(-i\bar{\Omega}_{\rm{NV}}-\Gamma_{\rm{NT}}\right)\hat{\sigma}_-+\Gamma_{\rm{NT}}\hat{\sigma}_+\nonumber \\
		&-\left[\left(i\bar{\Lambda}_{\rm{NT}}+\eta_{\rm{NT}}\frac{\Gamma_{\rm{NT}}}{2}\right)\hat{\sigma}_+^T+\left(i\bar{\Lambda}_{\rm{NT}}-\eta_{\rm{NT}}\frac{\Gamma_{\rm{NT}}}{2}\right)\hat{\sigma}_-^T\right]\hat{\sigma}_z,\\
		\partial_t\hat{\sigma}_-^T&=\left(-i\bar{\Omega}_{\rm{Tr}}-\eta_{\rm{NT}}^2\frac{\Gamma_{\rm{NT}}}{2}\right)\hat{\sigma}_-^T \nonumber\\
		&-\left[\left(i\bar{\Lambda}_{\rm{NT}}-\eta_{\rm{NT}}\frac{\Gamma_{\rm{NT}}}{2}\right)\hat{\sigma}_++\left(i\bar{\Lambda}_{\rm{NT}}-\eta_{\rm{NT}}\frac{\Gamma_{\rm{NT}}}{2}\right)\hat{\sigma}_-\right]\hat{\sigma}_z^T.
	\end{align}
	\label{QLESNT}
\end{subequations}
Eq.~(\ref{QLESNT}) illustrates that the dynamics between NV centers and transmon qubits are mutually influential but asymmetric, enabling the realization of nonreciprocal responses between NV centers and transmon qubits.
Figures~\ref{PRRFig12}(a) and (b) illustrate state conversion between the two, corresponding to the excitation of the NV center and the transmon qubit, respectively.
State conversion between the NV center and the transmon qubit is demonstrably nonreciprocal, as the NV center has difficulty transferring quantum states to the transmon qubit, whereas the transmon qubit can efficiently transfer quantum states to the NV center.

\section{\label{EF}Experimental feasibility}
In this setup, the disk has a radius $R=100~{\rm{nm}}$ and a thickness $h_G=5~{\rm{nm}}$, respectively, satisfying the condition $R\gg h_G$, which corresponds to a thin disk~\cite{2015BuettnerP225228,2021PaikarayP167900167900,2021SatywaliP19091909,2015DuP85048504,2013SunP167201167201,2015GilbertP84628462}.
The distance between the NV center and the disk surface is $d_G=5~{\rm{nm}}$.
Consequently, the coupling strength between them is $\Lambda_{\rm{SN}}/2\pi\approx12.5~{\rm{MHz}}$.
In contrast to the NV center, the transmon qubit is situated beneath the disk with the center coordinates of the SQUID are $x_c=R/2$, $y_c=0$, and $z_c=2 h_G$.
The diameter of the SQUID is taken as $100~{\rm{nm}}$.
Typical experimentally feasible parameters for transmon qubits include an externally applied flux $\phi_b=\pi/2$ and a Josephson energy imbalance of $\alpha_J=0.06$~\cite{2020KjaergaardP369395,2021BlaisP2500525005}.
The transverse coupling strength are calculated to be $\Lambda_{\rm{ST}}^T/2\pi\approx5.05~{\rm{MHz}}$.

Next, we analyze the strengths of coherent and dissipative coupling between the NV center and the transmon qubit, considering the intrinsic dissipation parameters of NV centers and transmon qubits as $\gamma_\mathrm{NV}^\mathrm{dc}/2\pi\approx 1~\mathrm{Hz}$, $\gamma_\mathrm{NV}^\mathrm{dp}/2\pi=10~\mathrm{kHz}$, $\gamma_\mathrm{Tr}^\mathrm{dc}/2\pi=20~\mathrm{kHz}$, and $\gamma_\mathrm{Tr}^\mathrm{dp}/2\pi=10~\mathrm{kHz}$~\cite{2020KjaergaardP369395,2020LiP153602153602,2013DohertyP145,2014RosskopfP147602147602,2013BarGillP17431743}.
For gyration-mode mediated coherent coupling, with gyration-mode dissipation $\gamma_\mathrm{GM}/2\pi=50~\mathrm{MHz}$~(see Sec.~\ref{DSGMMS}), the gyration-mode mediated indirect coupling $\Lambda_\mathrm{NT}/2\pi=0.5~\mathrm{MHz}$ between the NV center and the transmon qubit is obtained, with effective decay rates for the NV center and the transmon qubit given by $\Gamma_\mathrm{NV}^\mathrm{dc}/2\pi=0.48~\mathrm{MHz}$ and $\Gamma_\mathrm{Tr}^\mathrm{dc}/2\pi=0.1~\mathrm{MHz}$, respectively.
In other words, the indirect coupling between NV centers and transmon qubits can enter the strong coupling regime.
For gyration mode-mediated dissipative coupling, with $\Omega_2=\Delta_\mathrm{Tr}=\Delta_\mathrm{GM}=0$, the dissipative coupling strength simplifies to $\Gamma_{\rm{NT}}=4\bar{\Lambda}_\mathrm{SN}^2/\gamma_{\rm{GM}}$.
In the case of large dissipation ($\gamma_\mathrm{GM}\gg \bar{\Lambda}_\mathrm{SN}$), we set $\gamma_\mathrm{GM}/2\pi=300~\mathrm{MHz}$, yielding a dissipative coupling strength of $\Gamma_{\rm{NT}}/2\pi=0.52~\mathrm{MHz}$, which satisfies $\Gamma_{\rm{NT}}>\gamma_{\mathrm{NV},\mathrm{Tr}}^{\mathrm{dc},\mathrm{dp}}$, indicating that the system can enter the strong coupling regime.

\section{\label{Con}Conclusion}
We present a three-body hybrid quantum system integrating skyrmions, nitrogen-vacancy (NV) centers, and transmon qubits. By utilizing the gyration mode as a quantum bus, we establish both coherent and dissipative couplings between the NV center and transmon qubit that can reach the strong coupling regime. This enables bidirectional quantum state transfer with reciprocity control--reciprocal conversion can be realized through coherent coupling, while nonreciprocal state transfer is achievable via dissipative coupling mechanisms. The proposed hybrid platform provides a potential for developing integrated quantum computing architectures and advancing quantum information processing technologies at the chip scale.

\begin{acknowledgments}
P.-B.~L. is supported by the National Natural Science Foundation of China under Grants No. W2411002 and No. 12375018.
X.-F.~P. is supported by the National Natural Science Foundation of China under Grants No. 124B2091.
\end{acknowledgments}

\appendix
\section{\label{EHOT}The effects of higher-order terms}

\begin{figure}
	\centering
	\includegraphics[width=0.48\textwidth]{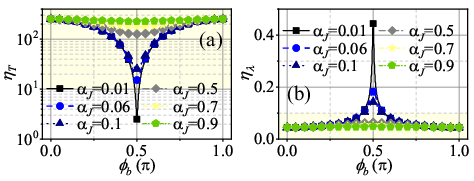}
	\caption{(a) and (b) display the variation of $\eta_T$ and $\eta_{\lambda}$ with the parameters $\alpha_J$ and $\phi_b$, respectively. The other parameters are $E_J^{\rm{max}}/h=50~{\rm{GHz}}$ and $E_C/h=200~{\rm{MHz}}$.}
	\label{PRRAPPFig1}
\end{figure}

In this section, we will analyze the effect of higher-order terms ($\hat{\widetilde{\delta}}^4$ and $\hat{\widetilde{\delta}}^3$) that were ignored in Sec.~\ref{SecHGT}.
The interaction Hamiltonian containing only higher-order terms is
\begin{equation}
	\hat{H}_{\rm{ST}}^\prime=\frac{E_J^{\rm{max}}\phi}{\mathcal{S}\left(\phi_b\right)}\left[\frac{\sin\left(2\phi_b\right)}{2}\frac{1-\alpha_J^2}{24}\hat{\widetilde{\delta}}^4+\frac{\alpha_J}{6}\hat{\widetilde{\delta}}^3\right].
	\label{HGTHOrder}
\end{equation}
Next, the first and second terms in Eq.~(\ref{HGTHOrder}) are calculated separately.
Substituting Eq.~(\ref{PhiQ2}) and Eq.~(\ref{QNDelta}) into
\begin{equation}
	\hat{H}_1^\prime=\frac{E_J^{\rm{max}}\phi}{\mathcal{S}\left(\phi_b\right)}\frac{\sin\left(2\phi_b\right)}{2}\left(1-\alpha_J^2\right)\left(\frac{1}{24}\hat{\widetilde{\delta}}^4\right)
\end{equation}
and simplifying it, we obtain
\begin{equation}
	\hat{H}_1^\prime={\Lambda_{\rm{ST}}^L}^\prime\left(\hat{a}+\hat{a}^\dagger\right)\left(\hat{b}^\dagger\hat{b}+\frac{1}{2}\hat{b}^\dagger\hat{b}^\dagger\hat{b}\hat{b}\right),
\end{equation}
where ${\Lambda_{\rm{ST}}^L}^\prime\equiv\eta_{\lambda}\Lambda_{\rm{ST}}^L$ and $\eta_{\lambda}=\sqrt{E_C/[2S(\phi_b)E_J^{\rm{max}}]}$.
Applying the same calculation, the second term in Eq.~(\ref{HGTHOrder}) can be reduced to
\begin{equation}
	\hat{H}_2^\prime={\Lambda_{\rm{ST}}^T}^\prime\left(\hat{b}^\dagger\hat{b}\hat{b}^\dagger\hat{a}+\hat{b}\hat{b}^\dagger\hat{b}\hat{a}^\dagger\right)
\end{equation}
with ${\Lambda_{\rm{ST}}^T}^\prime\equiv\eta_{\lambda}\Lambda_{\rm{ST}}^T$.
Here, the ground state $\vert g\rangle$ and the first excited state $\vert e\rangle$ are treated as qubits, then we can get $\hat{b}\rightarrow\hat{\sigma}_-^T$ and $\hat{b}^\dagger\rightarrow\hat{\sigma}_+^T$.
Utilizing $\hat{\sigma}_+^T\hat{\sigma}_+^T\hat{\sigma}_-^T\hat{\sigma}_-^T=0$, $\hat{\sigma}_+^T\hat{\sigma}_-^T\hat{\sigma}_+^T=\hat{\sigma}_+^T$, and $\hat{\sigma}_-^T\hat{\sigma}_+^T\hat{\sigma}_-^T=\hat{\sigma}_-^T$, the correction term of the interaction Hamiltonian due to the higher-order terms can be simplified to
\begin{equation}
	\hat{H}_{\rm{ST}}^\prime=\eta_{\lambda}\Lambda_{\rm{ST}}^L\left(\hat{a}+\hat{a}^\dagger\right)\hat{\sigma}_+^T\hat{\sigma}_-^T+\eta_{\lambda}\Lambda_{\rm{ST}}^T\left(\hat{a}\hat{\sigma}_+^T+\hat{a}^\dagger\hat{\sigma}_-^T\right).
\end{equation}
The interaction Hamiltonian, including the correction term, is
\begin{equation}
	\begin{split}
		\hat{H}_{\rm{ST}}^{\rm{corr}}&=-\Lambda_{\rm{ST}}^L\left(1-\eta_{\lambda}\right)\left(\hat{a}+\hat{a}^\dagger\right)\hat{\sigma}_+^T\hat{\sigma}_-^T\\
		&-\Lambda_{\rm{ST}}^T\left(1-\eta_{\lambda}\right)\left(\hat{a}\hat{\sigma}_+^T+\hat{a}^\dagger\hat{\sigma}_-^T\right).
	\end{split}
\end{equation}
This shows that the corrections introduced by the higher-order terms result in a reduction of the coupling strength.
We define the parameter $\eta_T=E_J^{\rm{max}}S(\phi_b)/E_C$ to describe the operational region of the transmon qubit, where $\eta_T\gg 1$ means the transmon qubit operates in the transmon regime.
Figure~\ref{PRRAPPFig1}(a) illustrates the variation of $\eta_T$ with $\phi_b$ and $\alpha_J$, where the shaded area represents the transmon regime.
It can be seen that superconducting qubits may operate beyond the transmon regime when $\alpha_J\rightarrow 0$ and $\phi_b\rightarrow \pi/2$.
In this case, the effect of higher-order terms will become significant [Fig.~\ref{PRRAPPFig1}(b)].
The shaded areas in Fig.~\ref{PRRAPPFig1}(b) represent regions where the correction terms can be safely ignored.
Here, we take $\alpha_J=0.06$ and $\phi_b=\pi/2$.
With these parameter values, the qubits operate in the transmon regime [Fig.~\ref{PRRAPPFig1}(a)].
However, in this case, the correction term contributes an effect greater than $0.1$, so the coupling strength after correction is  $\Lambda_{\rm{ST}}^{\rm{Tcorr}}/2\pi=\Lambda_{\rm{ST}}^T(1-\eta_{\lambda})=4.14~{\rm{MHz}}$.

\section{\label{SMCCNT}Skyrmion-mediated coherent coupling between an NV center and a transmon qubit}
Based on the previous analysis, the Hamiltonian of a hybrid quantum system comprising a skyrmion, an NV center, and a transmon qubit is given by Eq.~(\ref{HNST}).
By transforming to a rotating frame with frequency $\omega_\mathrm{Tr}$, the Hamiltonian takes the form  $\hat{H}_\mathrm{NST}=\Delta_\mathrm{GM}\hat{a}^\dagger\hat{a}+\Delta_\mathrm{NV}/2\hat{\sigma}_z+\Lambda_\mathrm{SN}(\hat{a}\hat{\sigma}_++\hat{a}^\dagger\hat{\sigma}_-)-\Lambda_\mathrm{ST}(\hat{a}\hat{\sigma}_+^T+\hat{a}^\dagger\hat{\sigma}_-^T)$, where the detunings are given by $\Delta_\mathrm{GM}=\omega_\mathrm{GM}-\omega_\mathrm{Tr}$ and $\Delta_\mathrm{NV}=\omega_\mathrm{NV}-\omega_\mathrm{Tr}$.
Next, we consider the adiabatic elimination of gyration modes under large detuning conditions.
By utilizing $\dot{\hat{O}}=i[\hat{H},\hat{O}]$, the equation of motion for the system's operator can be expressed as
\begin{subequations}
	\begin{align}
		\dot{\hat{a}}&=-\left(i\Delta_\mathrm{GM}+\frac{\gamma_\mathrm{GM}}{2}\right)\hat{a}-i\Lambda_\mathrm{SN}\hat{\sigma}_-+i\Lambda_\mathrm{ST}\hat{\sigma}_-^T, \label{AEEQOne}\\
		\dot{\hat{\sigma}}_-&=-\left(i\Delta_\mathrm{NV}+\frac{\gamma_\mathrm{NV}^\mathrm{dc}}{2}\right)\hat{\sigma}_-+i\Lambda_\mathrm{SN}\hat{a}\hat{\sigma}_z,\label{AEEQTwo}\\
		\dot{\hat{\sigma}}_-^T&=-\frac{\gamma_\mathrm{Tr}^\mathrm{dc}}{2}\hat{\sigma}_-^T-i\Lambda_\mathrm{ST}\hat{a}\hat{\sigma}_z^T.\label{AEEQThree}
	\end{align}
	\label{AEEQ}
\end{subequations}
Here, we examine the large-detuning regime with $\Delta_\mathrm{GM}\gg \Lambda_\mathrm{SN},\Lambda_\mathrm{ST},\gamma_\mathrm{GM}$.
Consequently, the formal integral of the equation of motion (\ref{AEEQ}) is given by
\begin{subequations}
	\begin{align}
		\hat{a}\left(t\right)&=\hat{a}\left(0\right)\exp\left(-\mathcal{X}_\mathrm{GM}t\right)+\exp\left(-\mathcal{X}_\mathrm{GM}t\right) \nonumber \\
		&\times\int_0^td\tau \left(-i\Lambda_\mathrm{SN}\hat{\sigma}_-+i\Lambda_\mathrm{ST}\hat{\sigma}_-^T\right)\exp\left(\mathcal{X}_\mathrm{GM}\tau\right), \label{FIEMOne} \\
		\hat{\sigma}_-\left(t\right)&=\hat{\sigma}_-\left(0\right)\exp\left(-\mathcal{X}_\mathrm{NV}t\right)+\exp\left(-\mathcal{X}_\mathrm{NV}t\right) \nonumber \\
		&\times\int_0^t d\tau\left(i\Lambda_\mathrm{SN}\hat{a}\hat{\sigma}_z\right)\exp\left(\mathcal{X}_\mathrm{NV}\tau\right),\\
		\hat{\sigma}_-^T\left(t\right)&=\hat{\sigma}_-^T\left(0\right)\exp\left(-\frac{\gamma_\mathrm{Tr}^\mathrm{dc}}{2}t\right)+\exp\left(-\frac{\gamma_\mathrm{Tr}^\mathrm{dc}}{2}t\right) \nonumber \\ &\times\int_0^td\tau\left(-i\Lambda_\mathrm{ST}\hat{a}\hat{\sigma}_z^T\right)\exp\left(\frac{\gamma_\mathrm{Tr}^\mathrm{dc}}{2}\tau\right),
	\end{align}
\end{subequations}
where $\mathcal{X}_\mathrm{GM}=i\Delta_\mathrm{GM}+\gamma_\mathrm{GM}/2$ and $\mathcal{X}_\mathrm{NV}=i\Delta_\mathrm{NV}+\gamma_\mathrm{NV}^\mathrm{dc}/2$.
We examine the regime in which the bosonic mode is significantly detuned.
Under this condition, the dynamics of the NV center and the transmon qubit are only weakly affected by the gyration mode.
Consequently, we can get
\begin{subequations}
	\begin{align}
		\hat{\sigma}_-\left(t\right)=&\hat{\sigma}_-\left(0\right)\exp\left[-\left(i\Delta_\mathrm{NV}+\frac{\gamma_\mathrm{NV}^\mathrm{dc}}{2}\right)t\right], \\
		\hat{\sigma}_-^T\left(t\right)=&\hat{\sigma}_-^T\left(0\right)\exp\left(-\frac{\gamma_\mathrm{Tr}^\mathrm{dc}}{2}t\right).
	\end{align}
	\label{PARTIALEQ}
\end{subequations}
Substituting Eq.~(\ref{PARTIALEQ}) into the integral equation of the gyration mode (\ref{FIEMOne})
and utilizing the condition $\gamma_\mathrm{GM}\gg \gamma_\mathrm{NV}^\mathrm{dc},\gamma_\mathrm{Tr}^\mathrm{dc}$, one can get
\begin{equation}
	\hat{a}\left(t\right)=\frac{-i\Lambda_\mathrm{SN}}{i\Delta_\mathrm{GM}+\gamma_\mathrm{GM}/2}\hat{\sigma}_-\left(t\right)+\frac{i\Lambda_\mathrm{ST}}{i\Delta_\mathrm{GM}+\gamma_\mathrm{GM}/2}\hat{\sigma}_-^T\left(t\right).
	\label{ATimeSim}
\end{equation}
Substituting Eq.~(\ref{ATimeSim}) into Eqs.~(\ref{AEEQTwo}) and (\ref{AEEQThree}), along with the condition $\hat{\sigma}_z\hat{\sigma}_-=-\hat{\sigma}_-$, yields the expression
\begin{subequations}
	\begin{align}
		\dot{\hat{\sigma}}_-&=-\left(i\Delta_\mathrm{NV}^\mathrm{eff}+\frac{\Gamma_\mathrm{NV}^\mathrm{dc}}{2}\right)\hat{\sigma}_-+i\Lambda_\mathrm{ST}\alpha\hat{\sigma}_z\hat{\sigma}_-^T,\\
		\dot{\hat{\sigma}}_-^T&=-\left(i\Delta_\mathrm{Tr}^\mathrm{eff}+\frac{\Gamma_\mathrm{Tr}^\mathrm{dc}}{2}\right)\hat{\sigma}_-^T+i\Lambda_\mathrm{ST}\alpha\hat{\sigma}_z^T\hat{\sigma}_-.
	\end{align}
\end{subequations}
Thus, the effective Hamiltonian describing the interaction between the NV center and the transmon qubit is given by
\begin{equation}
	\hat{H}_\mathrm{NT}^\mathrm{coh}=\frac{\Delta_\mathrm{NV}^\mathrm{eff}}{2}\hat{\sigma}_z+\frac{\Delta_\mathrm{Tr}^\mathrm{eff}}{2}\hat{\sigma}_z^T+\Lambda_\mathrm{NT}\left(\hat{\sigma}_+\hat{\sigma}_-^T+\hat{\sigma}_+^T\hat{\sigma}_-\right).
\end{equation}
Here, we have defined the effective parameters as $\Delta_\mathrm{NV}^\mathrm{eff}=\Delta_\mathrm{NV}-\alpha^2\Delta_\mathrm{GM}$, $\Delta_\mathrm{Tr}^\mathrm{eff}=-\beta^2\Delta_\mathrm{GM}$, $\Gamma_\mathrm{NV}^\mathrm{dc}=\gamma_\mathrm{NV}^\mathrm{dc}+\alpha^2\gamma_\mathrm{GM}$, $\Gamma_\mathrm{Tr}^\mathrm{dc}=\gamma_\mathrm{Tr}^\mathrm{dc}+\beta^2\gamma_\mathrm{GM}$, $\Lambda_\mathrm{SN}=\Lambda_\mathrm{ST}\alpha$, $\alpha\approx\Lambda_\mathrm{SN}/\vert\Delta_\mathrm{GM}\vert$, and $\beta\approx\Lambda_\mathrm{ST}/\vert\Delta_\mathrm{GM}\vert$.
Since the large detuning condition is considered, we can get $\alpha,\beta\ll 1$.
Returning to the original representation, we obtain the effective Hamiltonian
\begin{equation}
	\hat{H}_\mathrm{NT}^\mathrm{coh}=\frac{\omega_\mathrm{NV}^\mathrm{eff}}{2}\hat{\sigma}_z+\frac{\omega_\mathrm{Tr}^\mathrm{eff}}{2}\hat{\sigma}_z^T+\Lambda_\mathrm{NT}\left(\hat{\sigma}_+\hat{\sigma}_-^T+\hat{\sigma}_+^T\hat{\sigma}_-\right).
\end{equation}
where $\omega_\mathrm{NV}^\mathrm{eff}=\omega_\mathrm{NV}-\alpha^2\Delta_\mathrm{GM}$ and $\omega_\mathrm{Tr}^\mathrm{eff}=\omega_\mathrm{Tr}-\beta^2\Delta_\mathrm{GM}$.

%

\end{document}